\documentclass[conference]{IEEEtrans}

\usepackage{amsmath}
\usepackage{tikz}
\usepackage{xspace}
\usepackage{subfig}
\usepackage{makecell}

\newcommand{\rev}[1]{\textcolor{black}{#1}}
\newcommand{\IGNORE}[1]{}

\begin{document}

\title{
Characterizing the COVID-19 Transmission in South Korea Using the KCDC Patient Data
}

\author{
    \IEEEauthorblockN{Anna Schmedding}
    \IEEEauthorblockA{William \& Mary\\
    akschmedding@email.wm.edu}
    \and
    \IEEEauthorblockN{Lishan Yang}
    \IEEEauthorblockA{William \& Mary\\
    lyang11@email.wm.edu}
    \and
    \IEEEauthorblockN{Riccardo Pinciroli}
    \IEEEauthorblockA{Gran Sasso Science Institute\\
    riccardo.pinciroli@gssi.it}
    \and
    \IEEEauthorblockN{Evgenia Smirni}
    \IEEEauthorblockA{William \& Mary\\
    esmirni@cs.wm.edu}
}

\maketitle

\begin{abstract}
As the COVID-19 outbreak evolves around the world, the World Health Organization (WHO) and its Member States have been heavily relying on staying at home and lock down measures to control the spread of the virus.
In the last months, various signs showed that the COVID-19 curve was flattening, but even the partial lifting of some containment measures (e.g., school closures and telecommuting) appear to favor a second wave of the disease.
The accurate evaluation of possible countermeasures and their well-timed revocation are therefore crucial to avoid future waves or reduce their duration.
In this paper, we analyze patient and route data of infected patients from January 20, 2020, to May 31, 2020, collected by the Korean Center for Disease Control \& Prevention (KCDC).
This data analysis helps us to characterize patient mobility patterns and then use this characterization to parameterize simulations to evaluate different \textit{what-if} scenarios. Although this is not a definitive model of how COVID-19 spreads in a population, its usefulness and flexibility are illustrated using real-world data for exploring virus spread under a variety of circumstances.
\end{abstract}

\IEEEpeerreviewmaketitle

% \keywords{COVID-19, SARS-CoV-2, Coronavirus, Data Analysis, Agent-Based Model (ABM), Geographic Information System (GIS), Simulation}

\maketitle

\section{Introduction}
\label{sec:intro}

The first human cases of COVID-19 were observed in Wuhan, China, at the end of December 2019~\cite{rothan2020epidemiology}.
Since then, COVID-19 has spread in 114 countries all over the world. On March 11, 2020, the WHO declares COVID-19 the first pandemic caused by a coronavirus~\cite{who2020pandemic}.
Face masks~\cite{feng2020rational}, social distancing~\cite{lewnard2020scientific}, and, in the worst cases, quarantine and (partial) lockdown measures~\cite{sjodin2020only,nussbaumer2020quarantine} are the most commonly adopted strategies to flatten the infection curve.
The above measures proved effective in slowing down and limiting the pandemic outbreak~\cite{saez2020effectiveness}, but new COVID-19 waves are expected if mitigation measures are lifted~\cite{matrajt2020evaluating,smetters2020stay,who2020lifting}.
Simulation and mathematical models can be used to evaluate the effectiveness of proposed mitigation actions in advance, such as restrictions to population movement.
Typically, the effectiveness of models is tied up to their parameterization.
Past work  uses synthetic~\cite{kim2020location} or interaction data limited to small areas (e.g., households)~\cite{bock2020mitigation}.

The data sets used in this paper contain data collected by the Korean Center of Disease Control \& Prevention (KCDC) and local governments that, from January 20, 2020, to May 31, 2020, monitored infected people and logged their movements by using CCTV, cellphones, and credit card transactions~\cite{kcdc2020covid19}.
The KCDC records patient movements in plain text (i.e., natural language) without any unified rule.
Researchers parsed these logs through automated code and rule-based methods to extract keywords that are then used with web mapping service APIs (e.g., Google Maps~\cite{googlemap}, Kakao Map~\cite{kakaomap}, or Naver Map~\cite{navermap}) to extract geographical coordinates (i.e., latitude and longitude) and other data.
Extracted data are shared and made publicly available~\cite{kim2020ds4c}.
% In this paper, we use data of COVID-19 patients (and their routes) \cite{kim2020ds4c} collected by the Korean Center of Disease Control \& Prevention (KCDC) to study the COVID-19 outbreak in Seoul, South Korea.
% \evg{Please add information on the data: what form does it have, the dates covered from/to.}
Although the KCDC data set is a valuable resource for studying the spread of COVID-19, it presents some limitations that are described in the following.
\begin{itemize}
    \item South Korea has a small number of COVID-19 cases (i.e., 24,027 on October 3, 2020) compared to other countries, and the last version of the KCDC data set contains data collected up to May 31, 2020 (the KCDC data set has not been updated since then).
    On May 31, approximately 11,500 COVID-19 cases were confirmed in South Korea~\cite{kcdc2020covid19,kim2020spatiotemporal}, but only the 35\% of them have been logged into the data set.
    \item There is route data information for only a portion of the patients. Patient movement has been logged only for the
    15\% of all confirmed cases by May 31.
    \item Data are not collected evenly in the whole country. Although a good amount of data are available for Seoul, Gyeongsangbuk-do, and Gangwon-do, very little information is provided for other provinces.
    \item Some locations visited by patients are not recorded in the data set due to privacy issues. For this reason, patient information and route data do not always coincide.
    For example, there are patients that infect each other even if their routes do not cross.
    This may happen when patients belong to the same household (locations where people live are rarely logged in the data set).
    \item Patient and route data may be incomplete (i.e., some attributes are missing, such as the type of locations visited by some patients) and require manual intervention before analyzing the data set.
\end{itemize}

Different strategies are adopted to address the above challenges.
If some attributes are missing, they are manually retrieved by using available data. For example, in the case of patient routes with missing location type (e.g., store, school, hospital), other attributes, such as geographical coordinates, are used to retrieve the visited location and identify its type.
Data sets with missing data (e.g., movements of only the 15\% of confirmed cases are logged) cannot be always made complete by looking for extra information.
% \evg{I do not understand the next sentence.}
% While incomplete data is fixed by manually retrieving missing information, incomplete data sets cannot be adjusted by looking for other data online.
In this paper, we advocate using available data to extract information on movement habits of people living in Seoul (e.g., daily travel speed and distance based on patient age and day of the week).
We feed this information to a patched version of GeoMason~\cite{sullivan2010geomason}, a tool that uses agent-based models (ABM) and geographic information systems (GIS) to study disease outbreaks (e.g., a cholera outbreak was studied using this tool in~\cite{crooks2014agent}).
This way, we simulate\footnote{A demo of the simulation can be found at \url{https://youtu.be/H3qYZ47O6wU}.} interactions of thousands of people in Seoul on roads and in buildings to investigate the COVID-19 outbreak in the largest metropolis of South Korea and evaluate different \textit{what-if} scenarios. This tool offers a flexible model based on real-world COVID-19 spread information that can be used to facilitate evaluation of different mitigation measures and different patient behaviors.

The rest of the paper is organized as follows.
Sections~\ref{sec:dataset} and~\ref{sec:workload} describe the data sets used in this paper and their analysis.
Section~\ref{sec:simulation} presents the patched version of GeoMason and the data used to simulate the COVID-19 outbreak in Seoul.
%In Section~\ref{sec:scenario} we present different \textit{what-if} scenarios  using the simulation.
Section~\ref{sec:related} discusses related work and Section~\ref{sec:conclusion} concludes the paper.

\section{The KCDC Data Set}
\label{sec:dataset}

In this section, the KCDC data sets  are described.
The PatientInfo and PatientRoute data sets contain information and routes of COVID-19 patients in Seoul, respectively.
%The data sets \cite{kim2020ds4c} used in this paper contain data collected by the KCDC and local governments from January 20, 2020, to May 31, 2020. They monitor infected people (PatientInfo) and log their movements (PatientRoute) using CCTV, cellphones, and credit card transactions.
The amount of data in each data set is shown in Table \ref{tab:num_entries}.
Number of (healthy and sick) people moving across Seoul districts are also provided  in the SeoulFloating set. This data has been collected using the SK Telecom Big Data Hub.

\begin{table}[t]
    \centering
    \caption{Number of (unique) entries of PatientInfo and PatientRoute, two of the three data sets used in this paper.}
    \label{tab:num_entries}
    \begin{tabular}{c|cc}
    \hline
         & PatientInfo & PatientRoute \\
    \hline
        Total entries & 4004 & 8092 \\
        Unique patients & 4004 & 1472 \\
        Unique locations & -- & 2992 \\
        Unknown location \textit{type} & -- & 2341 \\
    \hline
    \end{tabular}
\end{table}

\noindent \textbf{PatientInfo data set.~}
This data set provides epidemiological data of COVID-19 patients. It contains 4004 different entries, each entry represents a different patient  identified by an ID (\textit{patient\_id}).
Other attributes include the gender and age (\textit{sex} and \textit{age}), their provenance (\textit{country}, \textit{province}, and \textit{city}), whether they have been infected in a known case (\textit{infection\_case}, e.g., overseas inflow or contact with patient) and the ID of the patient that infected them (\textit{infected\_by}), the number of people that the patient came in contact with (\textit{contact\_number}), and the date of their first symptoms (\textit{symptom\_onset\_date}).

\noindent \textbf{PatientRoute data set.~}
This data set contains 8092 entries, each one reporting a visit (to one of 2992 unique locations) of 1472 unique South Korean COVID-19 patients logged in this data set. A location is unequivocally identified by its \textit{latitude} and \textit{longitude}.
\textit{Province}, \textit{city}, and \textit{type} (e.g., airport, hospital, store) of each location are also provided.
Since the attribute \textit{type} of almost the 
30\% of entries is set to \textit{etc} (i.e., locations that cannot be identified using the rule-based approach of~\cite{kim2020ds4c}), we manually look for their type using their geographical coordinates and OpenStreetMap~\cite{openstreetmap}.
Each entry also contains the patient (\textit{patient\_id} (same as in the PatientInfo data set) and \textit{global\_num}, another ID used only in this data set) that visited the location on a specific \textit{date}.
The time spent on the location is not available. However, locations visited by a patient in a single day are logged in chronological order.

\noindent \textbf{SeoulFloating data set.~}
This data set provides hourly data of people moving across Seoul districts.
Data are collected from January 1 to April 30, 2020, by SK Telecom, a Korean wireless telecommunications operator.
Collected data are grouped by \textit{gender}, \textit{age}, and \textit{district} and allows visualizing the movement of people in Seoul during this period. Age is provided at the decade granularity for people in their 20s through 70s. No information is provided for children or for people who are 80 or more years old. Note that this data set reports data on the {\em entire} Seoul population, not just the COVID-19 patients, and only considers those who have cell phones.
\section{Workload Characterization}
\label{sec:workload}

Although the information contained in the considered data sets is not as accurate as one would like, it still allows for analyzing patient movements and interactions with high accuracy.
In this section, we discuss information that we extract from the analyzed data sets and how it is used as input in the GeoMason simulation tool~\cite{sullivan2010geomason}.

\subsection{Visited Locations}

\begin{figure}[t]
    \centering    
        \subfloat[South Korea. Blue points indicate hotspots.]{\label{subfig:south_korea}\includegraphics[width=0.32\columnwidth]{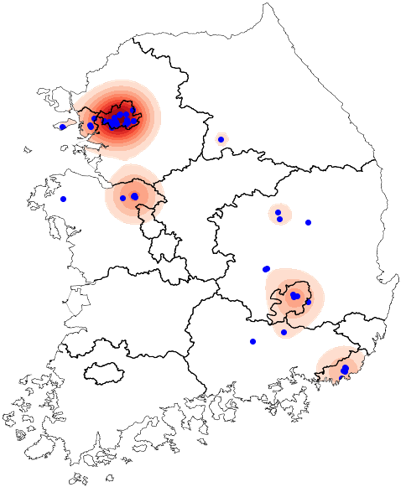}}
        \hfill
        \subfloat[Seoul. Gangnam district is outlined in blue.]{\label{subfig:seoul}\includegraphics[width=0.5\columnwidth]{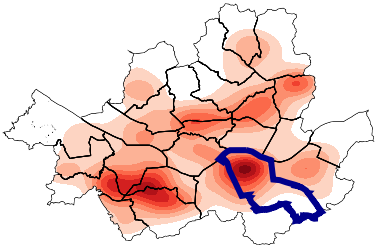}}
    \caption{Heat maps of most visited locations.}
    \label{fig:hotspot}
\end{figure}

Figs.~\ref{fig:hotspot}\subref{subfig:south_korea} and~\ref{fig:hotspot}\subref{subfig:seoul} depict a heat map of the most visited locations in South Korea and Seoul, respectively, showing  where COVID-19 outbreaks are more likely to happen.
Heat maps in Fig.~\ref{fig:hotspot} also show the South Korean cities for which movement data are recorded.
Visibly, Seoul is the city with the most visited locations. Within Seoul, the south-west and south-east areas are those with more patient routes.
The financial district and company head-quarters are located in the south-west part of the city.
The south-east region corresponds to the Gangnam district, see the district outlined in blue in Fig.~\ref{fig:hotspot}\subref{subfig:seoul}. Many shopping and entertainment centers are located in Gangnam.

\subsection{Seoul Population}

\begin{figure*}[htb]
    \centering
    \includegraphics[width=\textwidth]{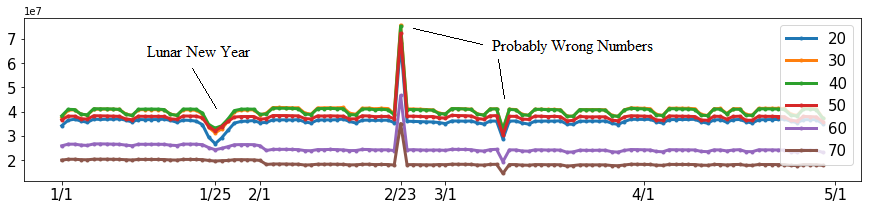}
    \caption{ Mobility of Seoul population over time by age group according to cell-phone data provided by SK telecom. }
    \label{fig:floating}
\end{figure*}

\begin{figure*}[htb]
    \centering    
        \subfloat[Patient connections (partial 1).]{\label{subfig:connection_partial1}\includegraphics[width=0.3\textwidth]{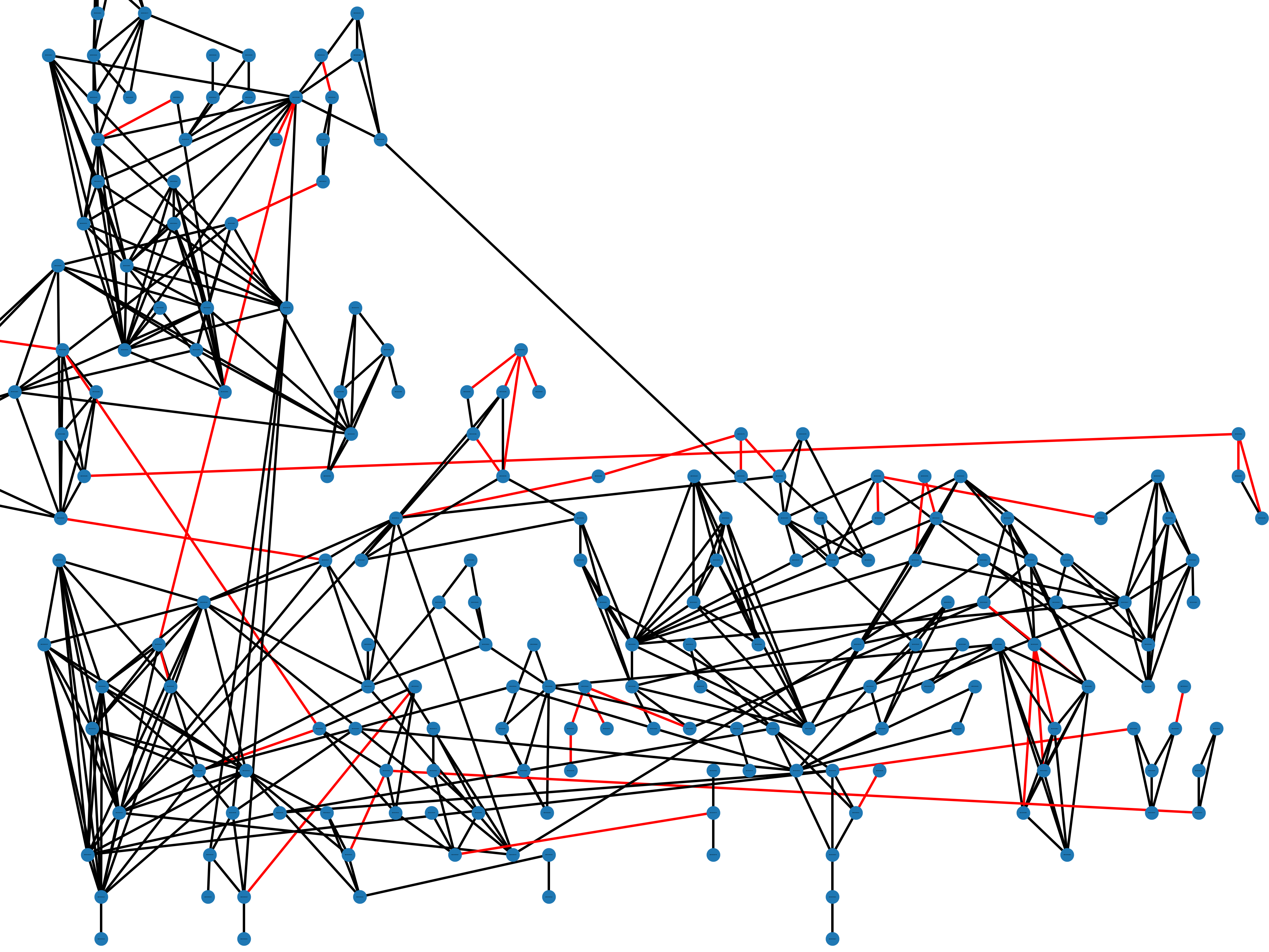}}
        \hfill
        \subfloat[Patient connections (partial 2).]{\label{subfig:connection_partial2}\includegraphics[width=0.3\textwidth]{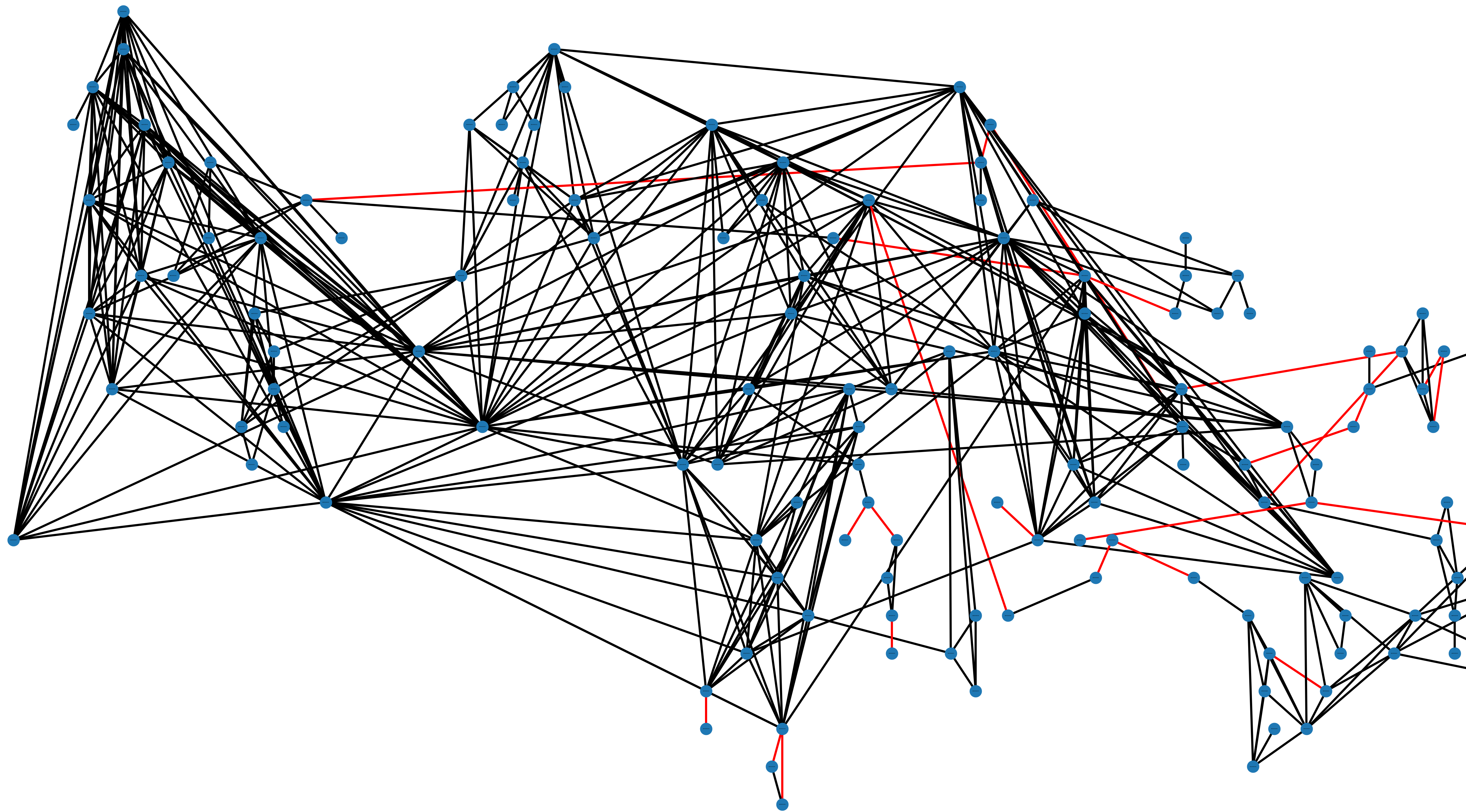}}
        \hfill
        \subfloat[Contact degree CDF.]{\label{subfig:contact_cdf}\includegraphics[width=0.3\textwidth]{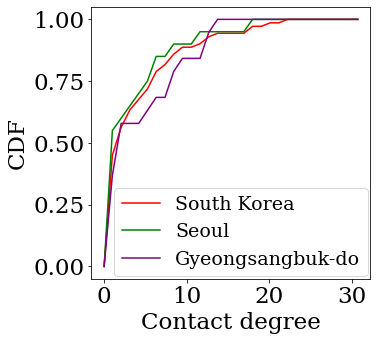}}
    \caption{Patient contacts.}
    \label{fig:contact-degree}
\end{figure*}

Since Seoul has more logs in PatientRoute as shown in Fig.~\ref{fig:hotspot}\subref{subfig:south_korea}, we analyze its population habits from January 1, 2020, to April 30, 2020, and extract information to determine how to put residents in different classes  to model  population movements.
Fig.~\ref{fig:floating} depicts the population (grouped by age) of both healthy and sick people moving in Seoul on a per-day basis.
Two clear classes of people are identified depending on their mobility: people that are 20 -- 50 years old (adults) and those that are 60 -- 70 (seniors).
The first group has higher mobility within the city during week days, but this mobility decreases during weekends.
The second group (seniors) does not have any discernible change in mobility patterns during the week.
A dip for the adult class observed on January 25 corresponds to the lunar new year day, no such dip is observed for the senior class.
The Seoul population shows a peak and a dip on February 23 and in March. Since there is no clear explanation regarding why, we treat these two days as measurement outliers and ignore them in the rest of the analysis.
Perhaps because of the pandemic onset in South Korean and KCDC advice, we observe the mobility of seniors to decrease at the beginning of February. 

Similar data is also available by splitting the dataset into men and women. Results are not present here due to lack of space and can be summarized as follows: there is no discernible difference between the mobility of men or women during the observed time period.

\subsection{Patient Connections}

Figs.~\ref{fig:contact-degree}\subref{subfig:connection_partial1} and~\ref{fig:contact-degree}\subref{subfig:connection_partial2} present
a subset of patient connections (to improve better visibility, we present here only a small portion of the graph of patient connections).
Here nodes depict patients, black edges connect patients that visited the same place during the same day, and red edges represent the virus spreading information obtained from the PatientInfo data set (i.e., \textit{infected\_by} attribute).
The node degree in Figs.~\ref{fig:contact-degree}\subref{subfig:connection_partial1} and~\ref{fig:contact-degree}\subref{subfig:connection_partial2} illustrates the contact degree among patients and illustrates visually the complexity of the problem.

Fig.~\ref{fig:contact-degree}\subref{subfig:contact_cdf} shows a summary view of the patient connections: the contact degree CDF of all patients for the entire dataset.
Specifically, three CDFs are shown: one for the whole South Korea, one for Seoul, and another one for Gyeongsangbuk-do (another Korean province).
Interestingly, all CDFs have a similar shape.

Figs.~\ref{fig:contact-degree}\subref{subfig:connection_partial1} and~\ref{fig:contact-degree}\subref{subfig:connection_partial2} show that some red edges do not overlap with black edges.
This means that, even if one of the two nodes connected by the red edge infected the other, no connections (i.e., visits to the same location during the same day) have been recorded in the data set.

\begin{figure*}
    \centering
    \includegraphics[width=\textwidth]{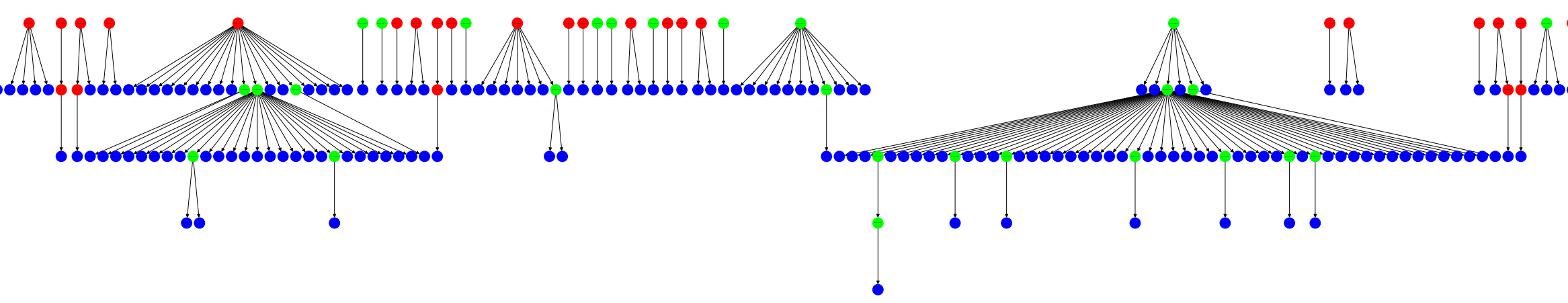}
    \caption{Subset of known infection spread between patients. Red nodes indicate patients with route information who infected others. Green nodes indicate patients who infected others but do not have any route information. Blue nodes indicate patients who did not infect anyone else.}
    \label{fig:bigfan}
\end{figure*}

\begin{figure*}[t]
    \centering    
        \subfloat[People infected.]{\label{subfig:spreader_infect}\includegraphics[width=0.25\textwidth]{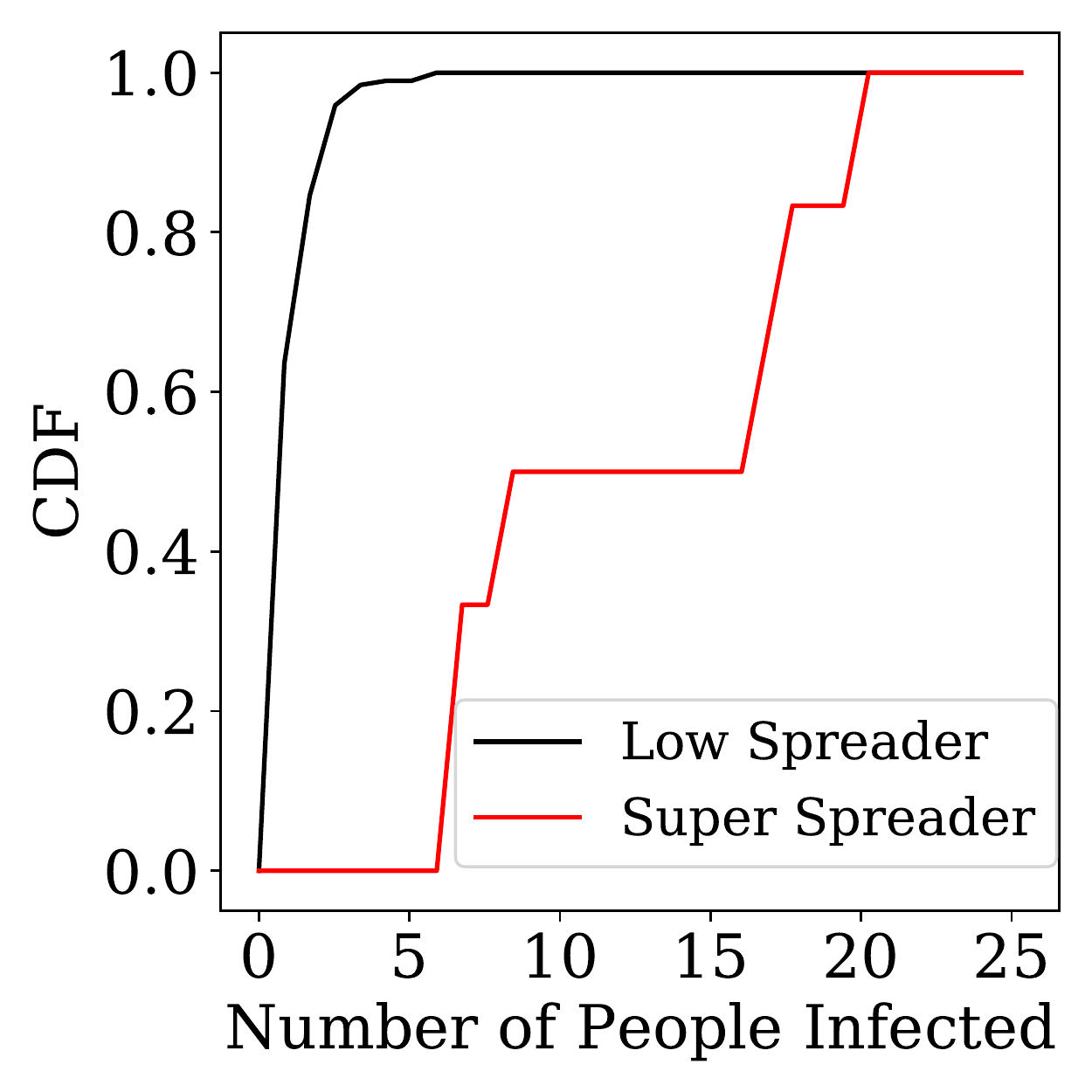}}
        \hfill
        \subfloat[Logged days.]{\label{subfig:spreader_date}\includegraphics[width=0.25\textwidth]{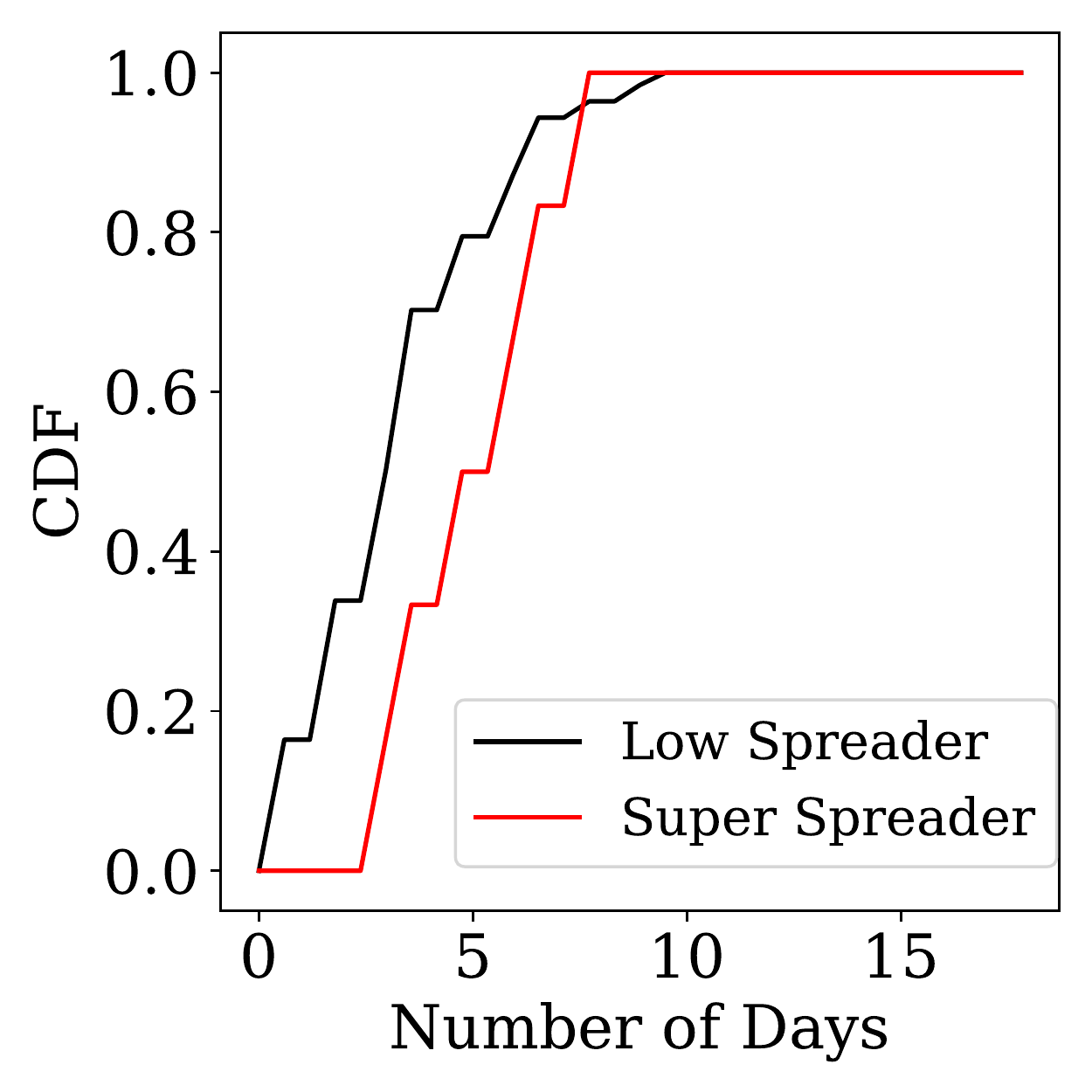}}    
        \subfloat[Unique locations visited.]{\label{subfig:spreader_location}\includegraphics[width=0.25\textwidth]{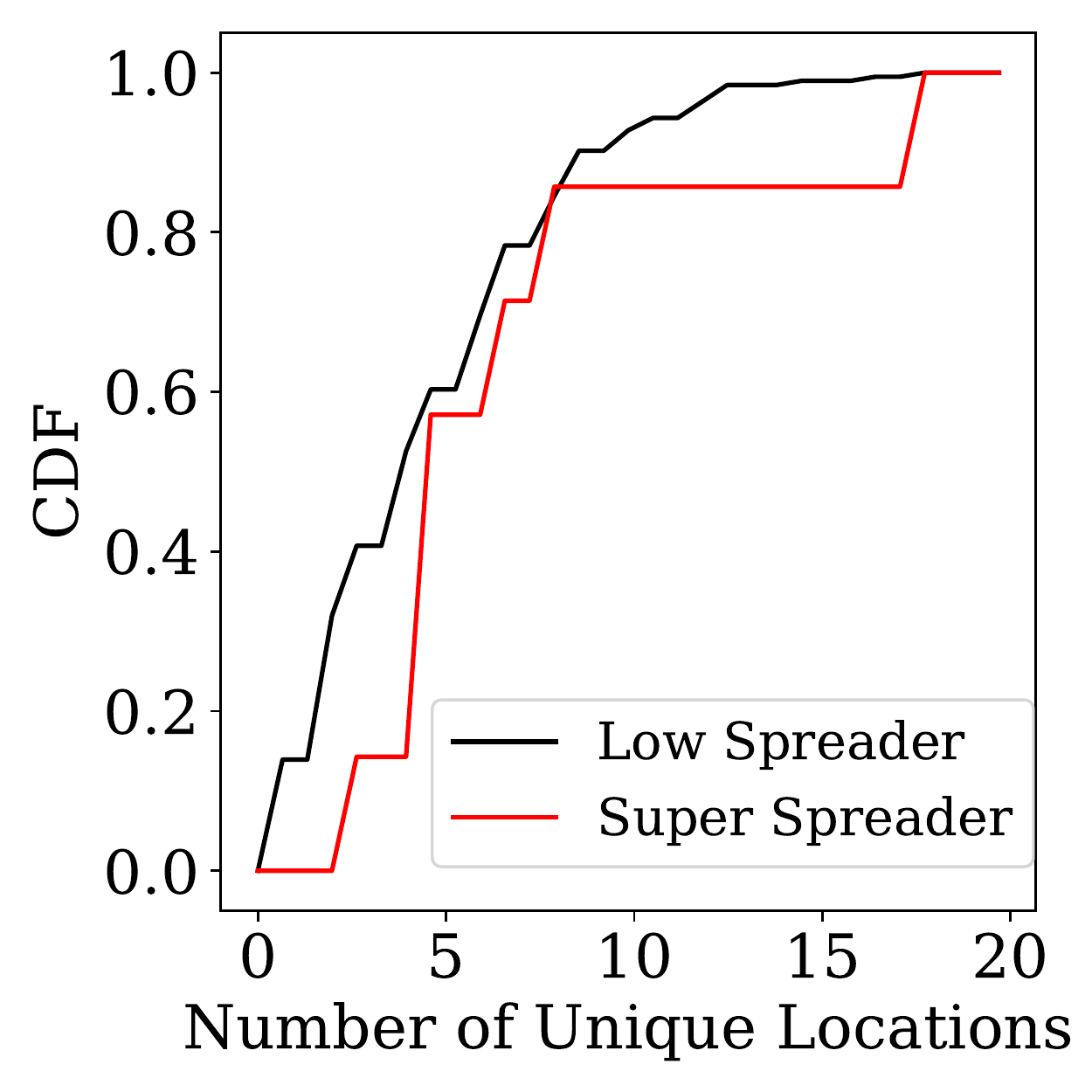}}
        \hfill
        \subfloat[Total locations visited.]{\label{subfig:spreader_records}\includegraphics[width=0.25\textwidth]{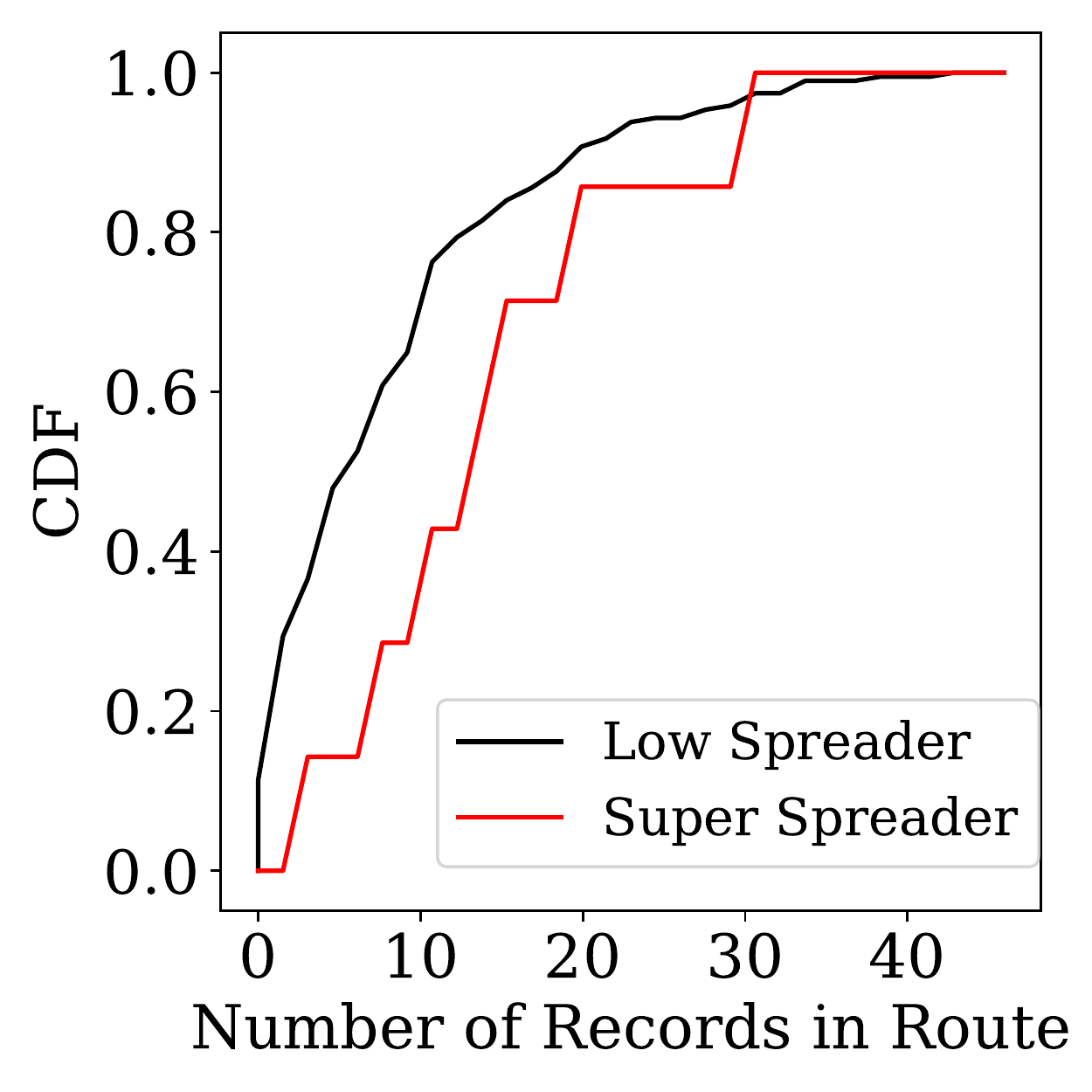}}
    \caption{Super spreader analysis.}
    \label{fig:superspreader}
\end{figure*}

\begin{figure}[htb]
    \centering
    \includegraphics[width=0.48\columnwidth]{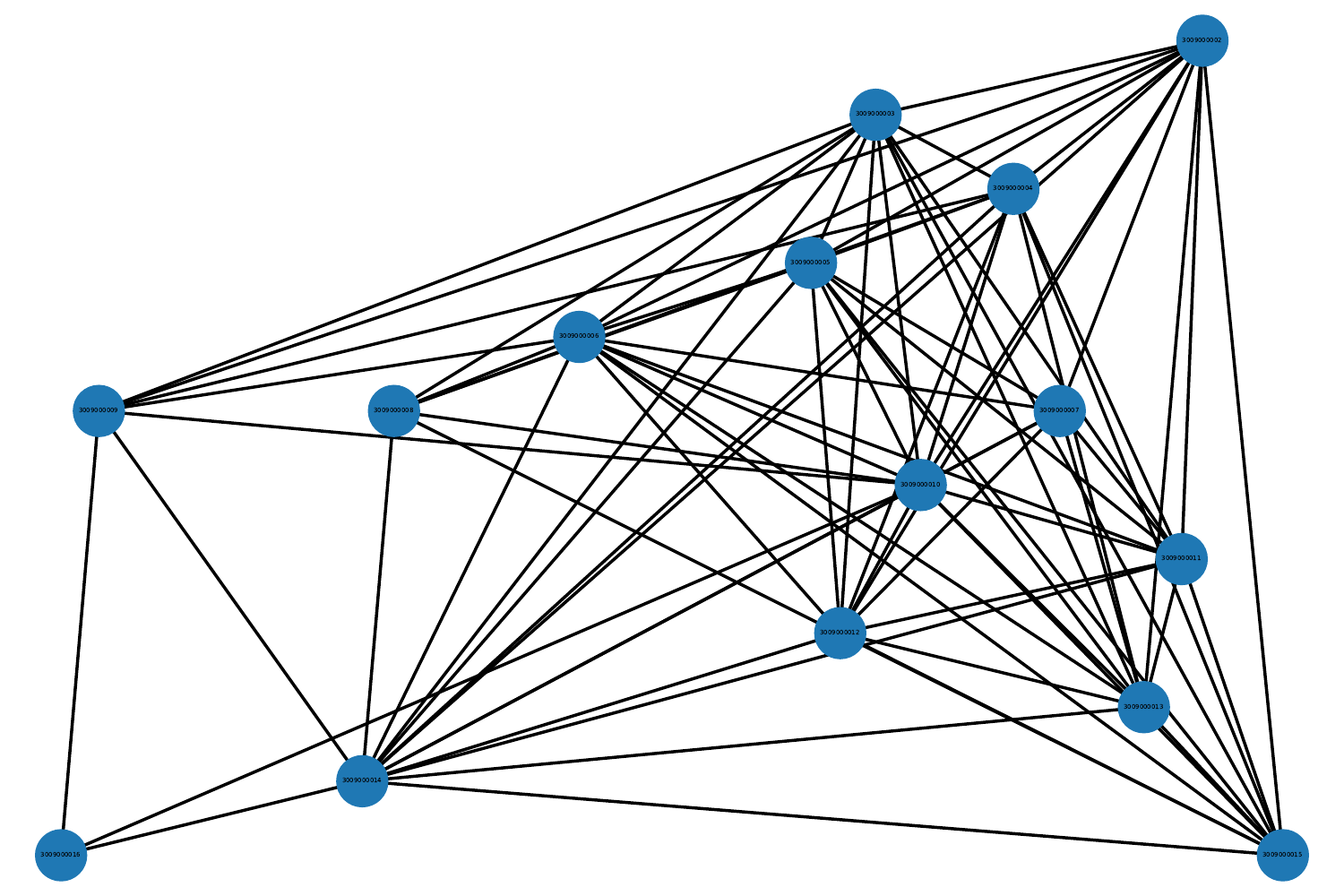}
    \caption{Outbreak associated with one particular hotspot location in Gangwon-do.}
    \label{fig:gangwondo_outbreak}
\end{figure}

High contact degree also indicates that super spreaders (i.e., patients that infect many other people) may exist.
However, people that came into contact with many others are not necessarily super spreaders since it is unknown whether or not they were sick or healthy when contact occurred. Because of this, further analysis is required to determine whether or not a patient is a super spreader.

\subsection{Super Spreaders}
\label{sec:superspreaders}

Fig.~\ref{fig:bigfan} illustrates a subset of patients where the {\em infected\_by} relationship (i.e., patient A is infected by patient B) is {\em known} from the PatientInfo data set. The entire graph contains 1052 patient nodes and 822 edges representing the known infection spread. 
Again, we present just a subset of the data for visibility here. Red nodes correspond to individuals with available route information who are known to have infected others, green nodes correspond to individuals who infected others but have no available route information and blue nodes correspond to patients who are not known to have infected others. 

This particular subset  shows a mix of super spreaders (i.e., people who infected more than six people) and low spreaders, who infected six or fewer people.\footnote{We define a ``super spreader" as someone who infected at least 6 people. This allows us to divide the data set to obtain the  most noticeable difference in patient behavior (number of locations, number of days, number of records).}
The large ``fans" in this figure are indicative of super spreaders.
The different behaviors of super spreaders and low spreaders are shown in Fig.~\ref{fig:superspreader}. Super spreaders account for 3.59\% and low spreaders account for the remaining 96.41\% of patients in this figure. 

Fig.~\ref{fig:superspreader} presents CDFs of the number of people infected by an individual, the number of days in the
log that the individual appears, the unique locations visited, and the total number of locations visited.
The CDFs in this figure indicate that, in general, super spreaders tend to be active for more days, visit more unique locations, and have longer routes than low spreaders. In particular, we see that all super spreaders in the data set are active for three or more days and visit three or more unique locations. Some of these super spreaders are active for up to 19 days and visit up to 18 unique locations with route lengths of up to 31 locations.

Outbreaks associated with specific locations (as opposed to specific super spreaders) can be examined individually as well. Fig.~\ref{fig:gangwondo_outbreak} depicts one specific outbreak at a law firm in the Gangwon-do province. Nodes indicate patients and edges indicate that patients were in the same location on the same date. The patients in the graph account for more than 150 visits to this location, causing repeated contact among individuals. This location is the biggest hot spot in the province.

\subsection{Daily Traveled Distance}

\begin{figure}[ht]
    \centering    
        \subfloat[Density heat map.]{\label{subfig:distance_density}\includegraphics[width=0.45\columnwidth]{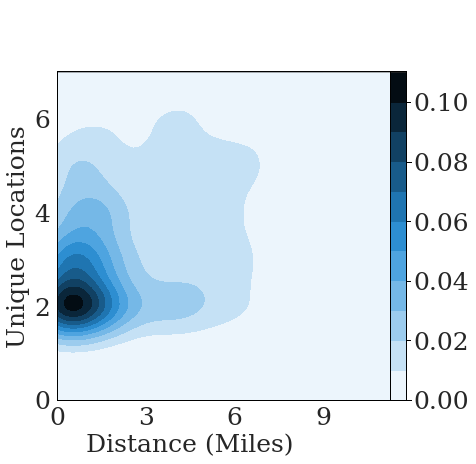}}
        \hfill
        \subfloat[Distance CDF.]{\label{subfig:distance_cdf}\includegraphics[width=0.4\columnwidth]{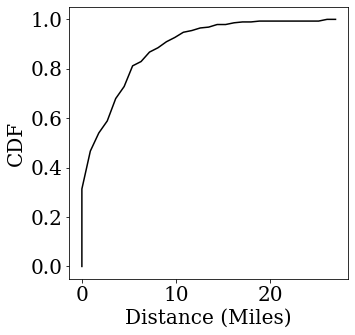}}
    \caption{Daily traveled distance and visited locations.}
    \label{fig:distance}
\end{figure}

Fig.~\ref{fig:distance}\subref{subfig:distance_density} plots the density heat map of distance traveled by patients in Seoul and the number of locations visited in a day, two important features due to the vital nature of patient movement to spread COVID-19.
The darker the area, the more patients have the same traveled distance and visited locations.
With some exceptions, people mostly travel short distances and visit only a few locations each day.
The CDF of the daily traveled distance is shown in Fig.~\ref{fig:distance}\subref{subfig:distance_cdf}.

\subsection{Patient Mobility}
\label{sec:mobility}

\begin{figure}[ht]
    \centering    
        % \subfloat[Patient count ]{\label{subfig:mobility_density}
        \includegraphics[width=0.46\columnwidth]{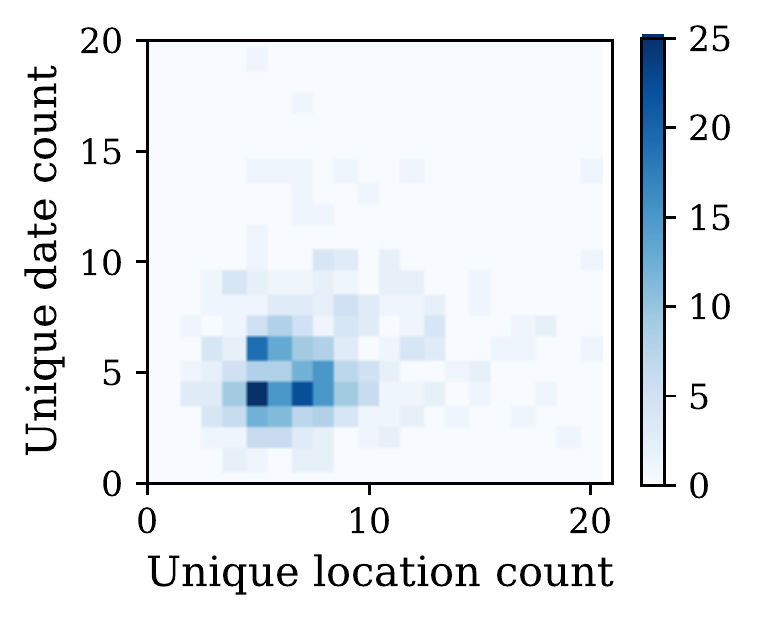}
    \caption{Patient count heatmap.}
    \label{fig:patient-count-heatmap}
\end{figure}

\begin{figure}[ht]
    \centering    
        \subfloat[Patient count ]{\label{subfig:location_count_hist}\includegraphics[width=0.42\columnwidth]{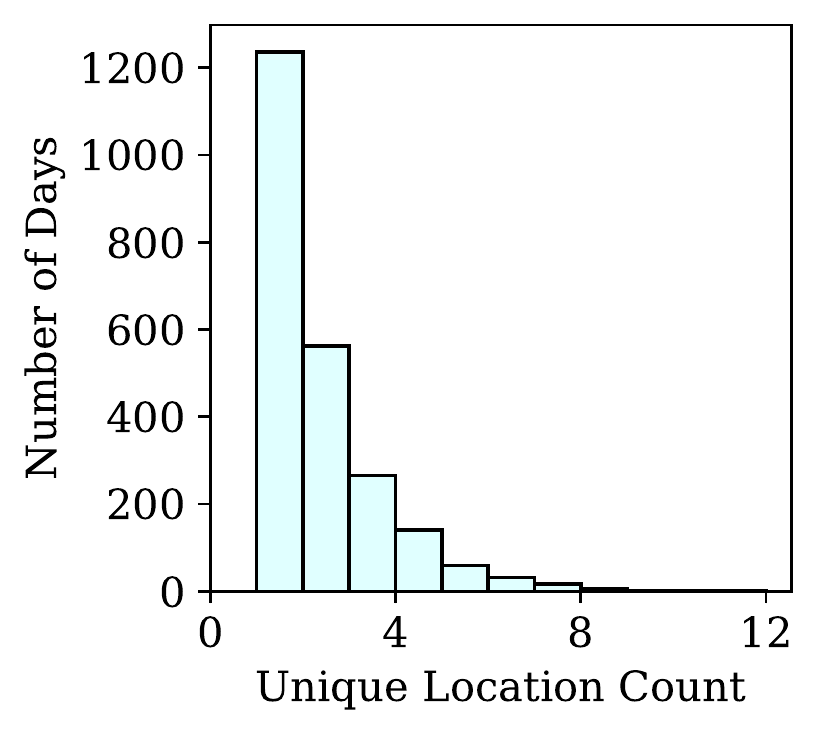}}
        \hfill
        \subfloat[Mobility CDF]{\label{subfig:mobility_cdf}\includegraphics[width=0.42\columnwidth]{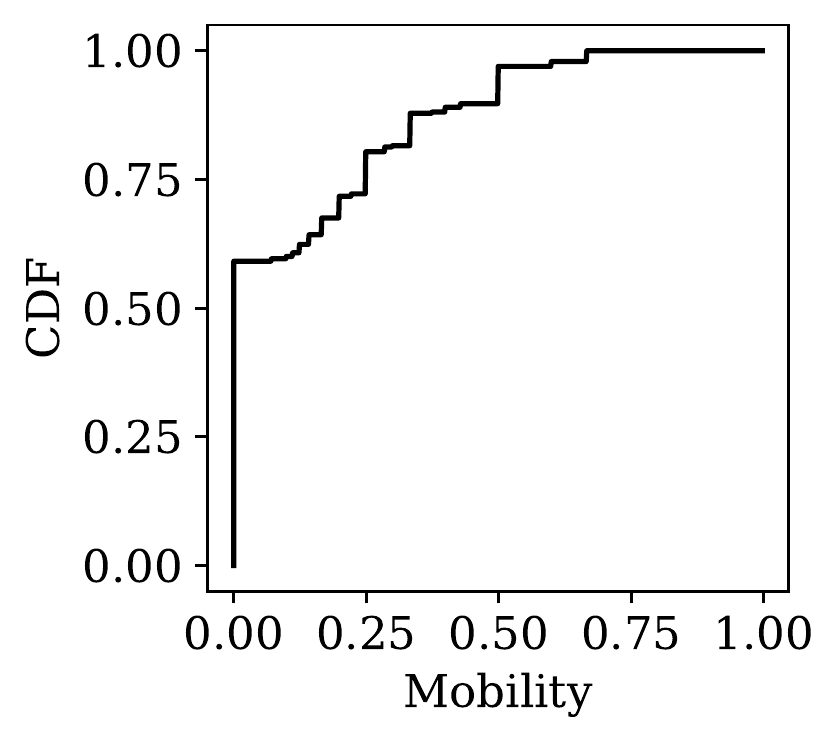}} \\
        \subfloat[Mobility CDF for low and super spreaders]{\label{subfig:mobility_spread}\includegraphics[width=0.42\columnwidth]{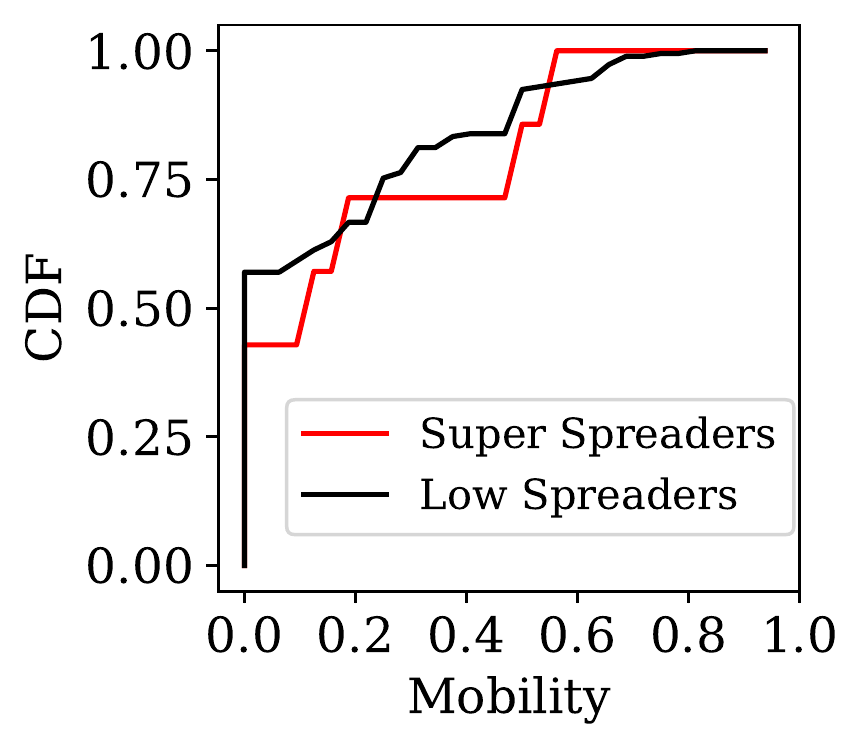}}
        \hfill
        \subfloat[Mobility CDF for young and senior patients]{\label{subfig:mobility_age}\includegraphics[width=0.42\columnwidth]{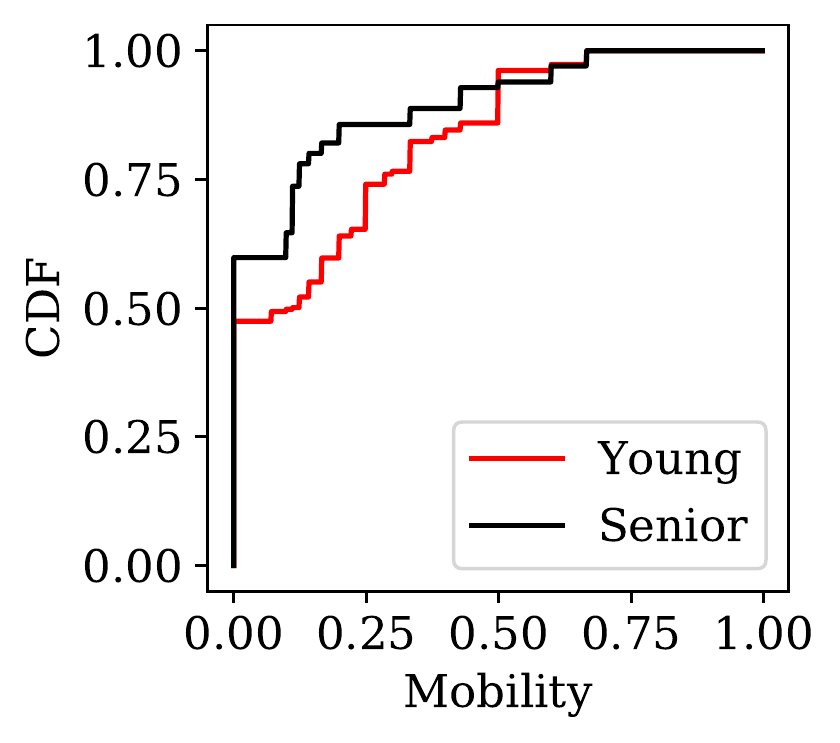}}
    \caption{Patient mobility.}
    \label{fig:mobility}
\end{figure}

Patient mobility is another important attribute that must be considered when studying the COVID-19 outbreak.
Intuitively, the more places a patient visits, the higher their mobility is.
 Fig.~\ref{fig:patient-count-heatmap} depicts the number of patients that are seen on 
a specific number of unique locations (x-axis) for a specific number of days (y-axis).
Note that the above graph does not distinguish patient mobility across {\em different} days. 
Indeed, looking at the mobility of individual patients, there are days where they exhibit high mobility and days where they move significantly less. This leads us to a more usable definition of mobility as a function of different time periods (days).
Fig.~\ref{fig:mobility}\subref{subfig:location_count_hist} shows the day count of unique locations reached by the patients in the data set: for
2,063  days (88.9\% of days) a typical patient visits 1--3 locations, while for 258 days  (11.1\%) more than 3 unique locations are visited.

% This requires the mobility of a patient to be defined by considering both their \textit{high} and \textit{low} mobility days.
Defining a \textit{high mobility day} as a day during which a patient visits at least $L$ locations, the {\em mobility of a patient} is defined as the ratio of the patient high mobility days to all logged days for this specific individual. Note that this is not the only way to define mobility. For simulation purposes (see Section~\ref{sec:simulation}), this definition provides a practical way to capture mobility with a probability.
Based on the histogram shown in Fig.~\ref{fig:mobility}\subref{subfig:location_count_hist}, days with $L \le 3$ are considered of low mobility.
The CDF of patient mobility using the above definition is depicted in Fig.~\ref{fig:mobility}\subref{subfig:mobility_cdf}. The figure shows that 57.6\% of patients never visit more than 4 locations in a day.

Different classes of patients have different mobility.
Fig.~\ref{fig:mobility}\subref{subfig:mobility_spread} shows the difference in mobility between super spreaders and low spreaders, while
Fig.~\ref{fig:mobility}\subref{subfig:mobility_age} illustrates mobility by age groups. 
Super spreaders and young people have higher mobility compared to low spreaders and seniors, respectively.
\rev{For higher percentiles, the low spreaders have larger mobility than super spreaders due to the small number of super spreader agents available in the KCDC data set.}

\subsection{Irresponsible Behaviors}
\label{sec:irresponsible}

\begin{figure}[htb]
    \centering    
        \subfloat[Active days after first symptoms.]{\label{subfig:onset_date}\includegraphics[height=0.32\columnwidth]{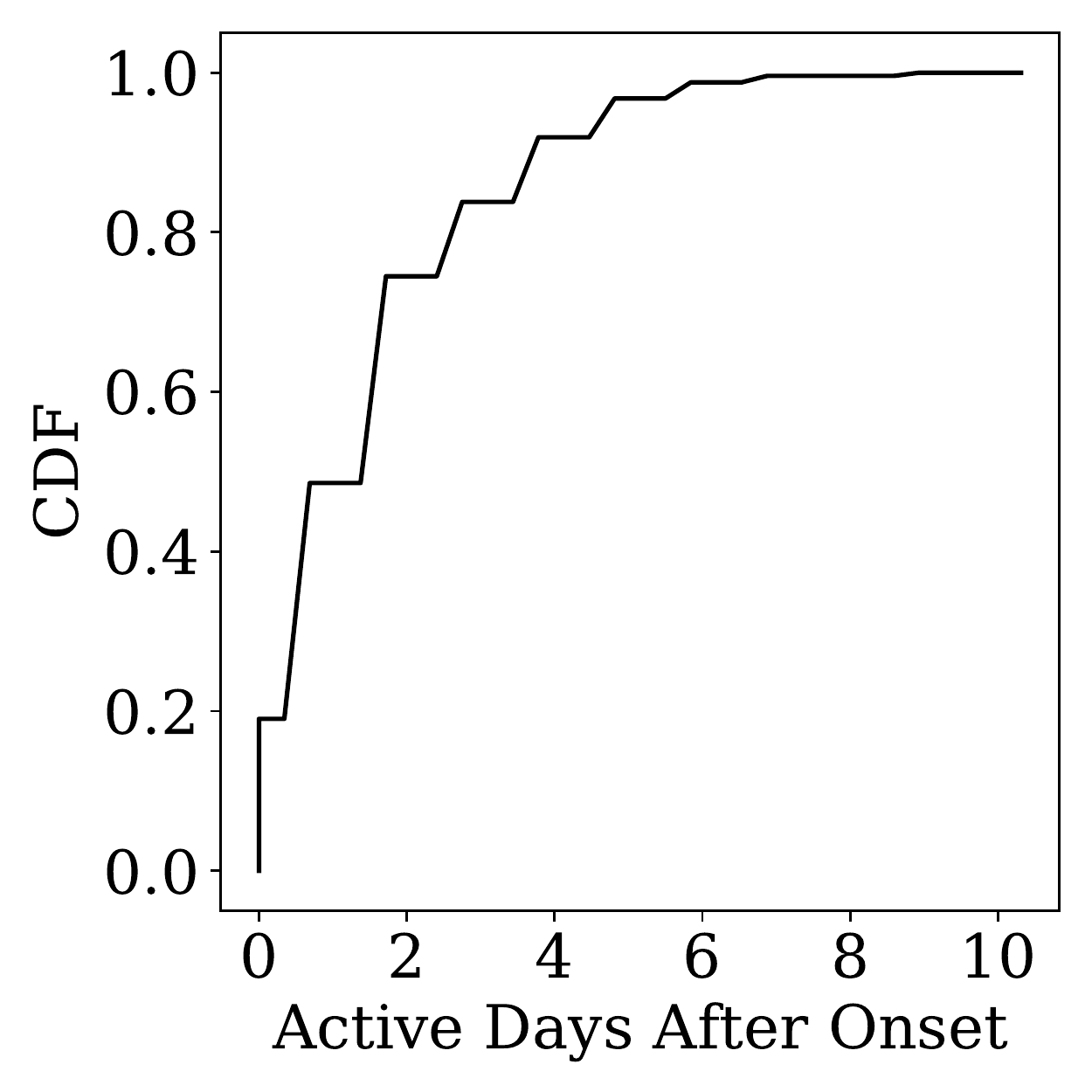}}
        \hfill
        \subfloat[Unique locations visited after first symptoms.]{\label{subfig:onset_location}\includegraphics[height=0.32\columnwidth]{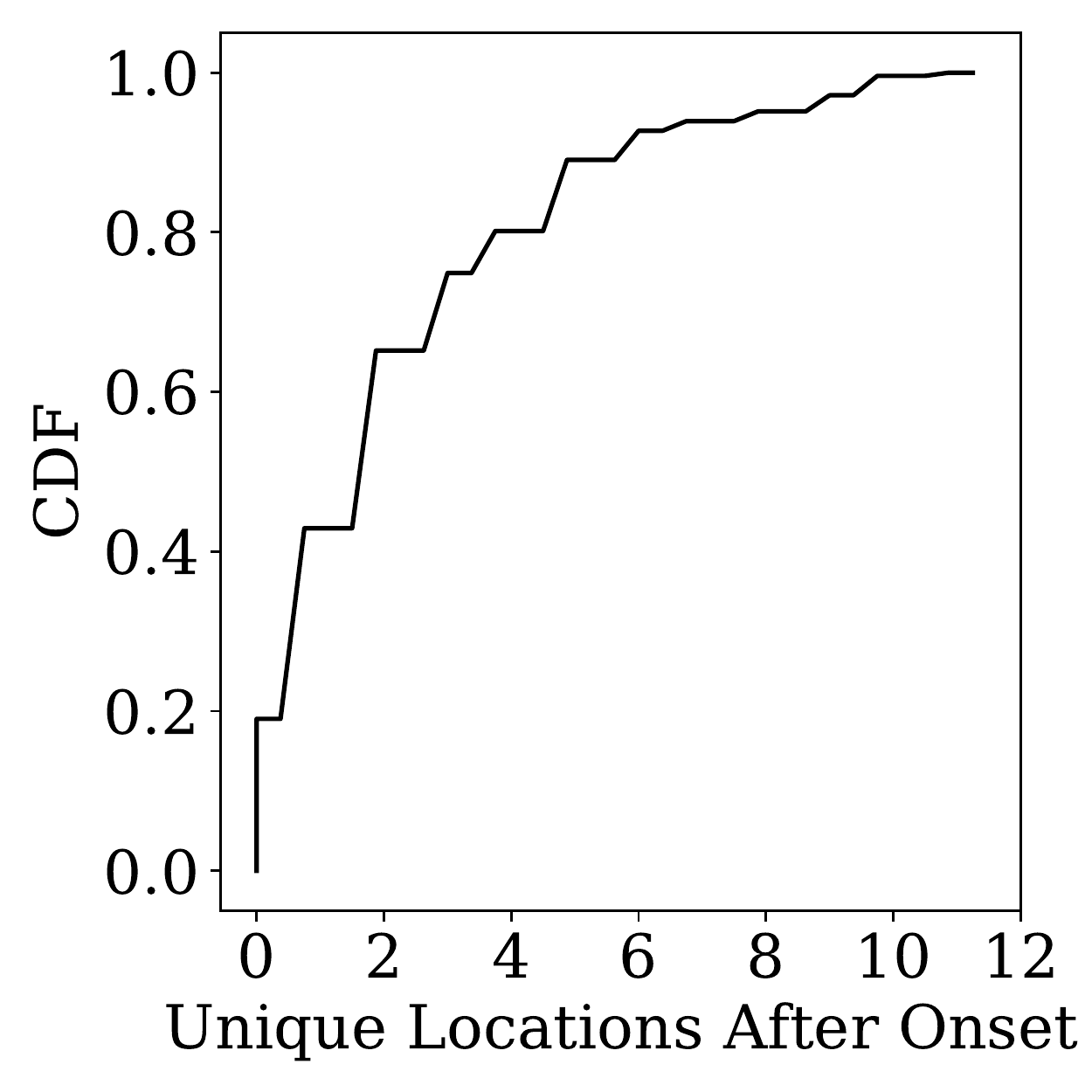}}  
        \hfill
        \subfloat[Total locations visited after first symptoms.]{\label{subfig:onset_record}\includegraphics[height=0.32\columnwidth]{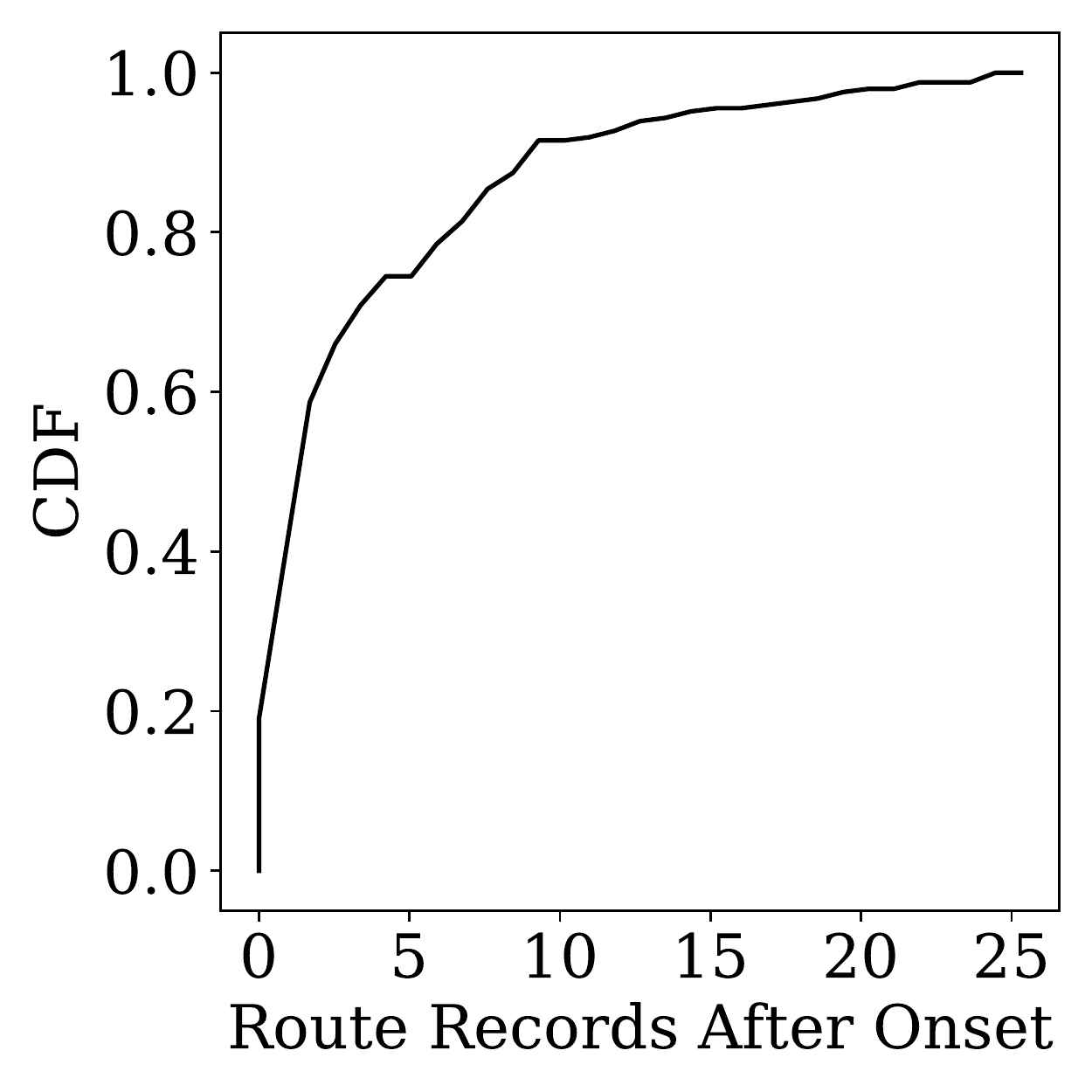}}
    \caption{Irresponsible behavior of sick patients: mobility after symptom onset. }
    \label{fig:onset}
\end{figure}

\begin{figure*}[!tb]
	\centering
	\begin{minipage}[t]{\textwidth}
		\centering
 		\includegraphics[scale=0.9]{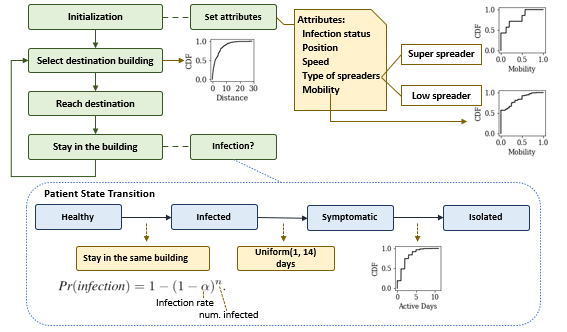}
	\end{minipage}
% 	\vspace{-\baselineskip}
% 	\vspace{-1em}
	\caption{Life cycle of an agent.}
	\label{fig:agent-life-cycle}
% 	\vspace{-1em}
\end{figure*}

Patients behave irresponsibly when they keep moving after the onset of their first COVID-19 symptoms.
This clearly facilitates the diffusion of the disease since sick people can still meet and infect healthy ones.
We present how long sick people continue to show mobility after exhibiting symptoms.
Results are shown in Fig.~\ref{fig:onset}, where one can see is that only the 20\% of patients stop moving and isolate immediately after initial symptoms are observed.
Some patients keep moving for more than a week after initial symptoms were observed, see Fig.~\ref{fig:onset}\subref{subfig:onset_date}.
They also visit many locations. Figs.~\ref{fig:onset}\subref{subfig:onset_location} and~\ref{fig:onset}\subref{subfig:onset_record} show the number of unique and total locations that sick patients visit after initial symptoms are observed.

% Another question we are interested in is how much patients move, i.e., the mobility of patients. 
% % To be able to define the mobility, we count the number of locations a patient visit during one day. 
% Intuitively, the more places a patient visits, the higher the mobility is. 
% The most active patient visited 21 places in a single day.
% However, when exploring the dataset, we found there are some patients who visited many places for some days, and only one or two places for other days.
% With the definition of a high mobility day as visiting at least 4 locations, the mobility of a patient is defined as the percentage of high mobility days.
% The CDF is shown on the right of Fig.~\ref{fig:agent-life-cycle}.
% Around 60\% of the patients are completely low mobility.  

\section{Simulation}
\label{sec:simulation}

In this section, we show how to parameterize a simulation based on a patched version of GeoMason~\cite{sullivan2010geomason} using the characterization presented in Section~\ref{sec:workload}.
The attributes, life cycle, and states of an agent are shown in Figure~\ref{fig:agent-life-cycle}.
The following attributes are set during the initialization phase:
\begin{enumerate}
    
    \item {\sl Infection status.} A random agent is selected as patient zero in the considered area. 
    \item {\sl Position.} Agents are randomly placed on a road in the simulated area.
    \item {\sl Speed.} There are two types of agents: 50\% of agents are considered pedestrian and walk at a speed of 3 MPH before reaching their destination; other agents drive a vehicle and their speed is uniformly distributed between 10 and 25 MPH.
    \item {\sl Type of spreaders.} We define two classes of spreaders: 3.59\% of patients are super spreaders and 96.41\% are low spreaders (see Section~\ref{sec:superspreaders}).
    \item  {\sl Mobility.} 
    We use the mobility of super spreaders and low spreaders depicted in  Fig.~\ref{fig:mobility}\subref{subfig:mobility_spread} to model different types of patient mobility.
\end{enumerate}
In addition to the mobility distribution of super spreaders and low spreaders, the CDF of daily traveled distance in Fig.~\ref{fig:distance}\subref{subfig:distance_density} is also used to determine the distance to a destination.

\begin{figure}[tb]
	\centering
	\begin{minipage}[t]{\columnwidth}
		\centering
 		\includegraphics[scale=0.26]{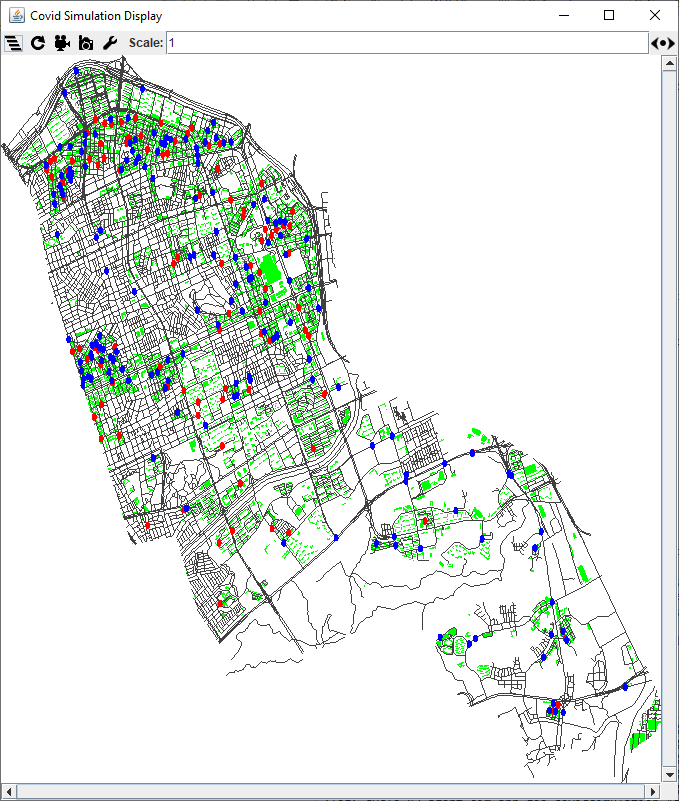}
	\end{minipage}
	\vspace{-\baselineskip}
% 	\vspace{-1em}
	\caption{A screenshot of the simulation: Gangnam district.}
	\label{fig:simulation-screenshot}
% 	\vspace{-1em}
\end{figure}

% \begin{figure}[tb]
% \vspace{-\baselineskip}
% 	\centering
% 	\begin{minipage}[t]{0.52\columnwidth}
% 		\centering
%  		\includegraphics[scale=0.47]{figs/infection_norm_5000.png}
 		
%  		\vspace{-0.5\baselineskip}
%  		(a) Population=5,000
% 	\end{minipage}
% 		\begin{minipage}[t]{0.47\columnwidth}
% 		\centering
%  		\includegraphics[scale=0.47]{}
 		
%  		\vspace{-0.5\baselineskip}
%  		(b) Population=10,000
% 	\end{minipage}
% 	\vspace{-1.1\baselineskip}
% 	\caption{
% % 	Simulation timeline in terms of different infected population.
% % 	Percentage of infected population grows under different simulation settings.
% 	CDF of infected agents within 3,000 minutes.}
% 	\label{fig:simulation-results}
% 	\vspace{-1.1\baselineskip}
% \end{figure}

The simulation time is defined by cycles. In each simulation cycle, agents outside a building move along the road towards their destination; agents inside a building can choose to stay or leave, based on their mobility. Agents with high mobility have a high probability to leave the building. 
Note that agents stay in a building for at least 15 minutes in order to meet the definition of close contact~\cite{CDCguidance}.
If multiple agents are inside the same building, they may infect each other with a certain probability.
When  infection happens, the agent state changes from healthy to infected, as the state transition shown in Fig.~\ref{fig:agent-life-cycle}.
We assume the outdoor infection probability to be negligible. Given the probability of infection inside a building, $\alpha$, and the number of infected agents in the building, $n$, the probability of a healthy agent  to be infected by a contact within the building is:
% \vspace{-1mm}
\begin{equation}
\label{eq:infection}
    % \vspace{-\baselineskip}
    \textit{Pr}(\textit{infection}) = 1 - (1-\alpha)^n.
    % \vspace{-1mm}
\end{equation}

It takes 1--14 days for patients to show symptoms after infection according to the WHO~\cite{whoQandA}.
We therefore use a Uniform distribution between 1 and 14 days to transition from infected to symptomatic.
Since there exist patients who continue to move even after showing symptoms, as seen in Fig.~\ref{fig:onset}, we 
 use the CDF in Fig.~\ref{fig:onset}\subref{subfig:onset_date} to determine the number of active days after their first symptoms.
We do not distinguish the behavior of super and low spreaders because of lack of data (there are only two super spreaders with  \textit{symptom\_onset} information available).
After each infected person exhausts their 
 active days after infection, they are isolated. %(self-quarantine or hospitalization). 

Consistent with infectious disease simulation studies~\cite{kim2020location}, we set the simulation cycle  to  5 minutes.
The simulation stops when all agents are infected.
We simulate the COVID-19 outbreak in the Gangnam district, i.e., the sub municipality of Seoul with the most hotspots, see Fig.~\ref{fig:hotspot}\subref{subfig:seoul}.  
This area has 11,438 road intersections and 7,043 buildings.
Roads and buildings are placed in the simulated area as described in \cite{seoulGIS}, a collection of GIS data with regard to Seoul.
GeoMason loads the  GIS data (e.g., roads, road intersections, buildings) stored in a shapefile format, i.e., a file that stores geometric locations and their attribute information.
Although the longest distance we observe in PatientRoute data set in Seoul is 30 miles, the longest distance between two buildings in the simulated Gangnam district is 7.06 miles. Therefore, we normalize the maximum distance to  3.53, which is half of the longest distance in the simulated area, to ensure a valid building selection as the agent's destination.  
Gangnam district district's population is 604,586 and a total of 7,043 buildings. We do not have any information on the building stories, entries, or number of rooms. This information is crucial, especially for apartment buildings, where multiple people can be inside the same building at the same time without contact. 
To address this lack of information, we limit the population in our simulations.

A screenshot of the GeoMason simulation execution can be seen in Fig.~\ref{fig:simulation-screenshot}: the black lines are roads that agents travel on, and green areas are buildings. 
Agents only have two states in terms of infection, i.e., infected (red dots) or healthy (blue dots).

\begin{figure}[tb]
    \centering   
    \includegraphics[scale=0.53]{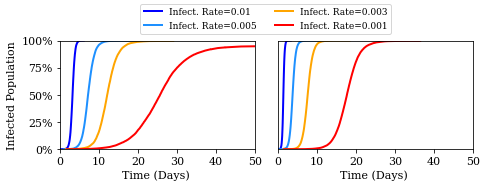}
    
    \begin{minipage}{0.55\columnwidth}
	\centering
	\footnotesize
	(a) Population = 10,000.
\end{minipage}
    \begin{minipage}{0.42\columnwidth}
	\centering
	\footnotesize
	(b) Population = 20,000.
\end{minipage}

        % \subfloat[Population = 10,000.]{\label{subfig:symptom-pop10k}\includegraphics[scale=0.45]{figs/final_pop10000.pdf}}
        % \hfill
        % \subfloat[Population = 20,000.]{\label{subfig:symptom-pop20k}\includegraphics[scale=0.45]{figs/final_pop20000.pdf}}
        
        % \subfloat[Population = 10,000.]{\label{subfig:symptom-pop10k}\includegraphics[scale=0.5]{figs/tmp_results/cdf_exp_covid_symptom_pop10000_rate0.01_dist1.pdf}}
        % \hfill
        % \subfloat[Population = 20,000.]{\label{subfig:symptom-pop20k}\includegraphics[scale=0.5]{figs/tmp_results/cdf_exp_covid_symptom_pop20000_rate0.01_dist1.pdf}}
    \caption{Percentage of infected population  when simulating patient isolation.}
    \label{fig:symptom-simulation-results}
\end{figure}

Fig.~\ref{fig:symptom-simulation-results} depicts the percentage of infected population as a function of time. The simulation stops at 50 days. 
The graph illustrates how quickly the entire population is infected for four infection rates that correspond to measures such as mask wearing and social distancing.
The figure includes results for two population sizes and shows the speed of the disease spread as a function of population density, infection in 
Fig.~\ref{fig:symptom-simulation-results}(b) is faster than Fig.~\ref{fig:symptom-simulation-results}(a).

\begin{figure}[tb]
    \centering    
    \includegraphics[scale=0.53]{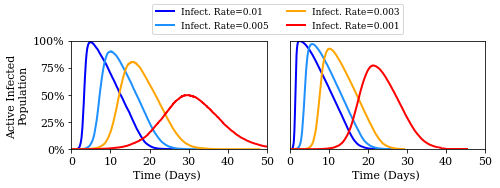}
    
        \begin{minipage}{0.55\columnwidth}
	\centering
	\footnotesize
	(a) Population = 10,000.
\end{minipage}
    \begin{minipage}{0.42\columnwidth}
	\centering
	\footnotesize
	(b) Population = 20,000.
\end{minipage}

        % \subfloat[Population = 10,000.]{\label{subfig:active-10k}\includegraphics[scale=0.43]{figs/active_pop10000.pdf}}
        % \hfill
        % \subfloat[Population = 20,000.]{\label{subfig:active-20k}\includegraphics[scale=0.43]{figs/active_pop20000.pdf}}
    \caption{Percentage of active while infected population.}
    \label{fig:symptom-simulation-results-active}
\end{figure}

\begin{figure}[!tb]
    \centering    
    \includegraphics[scale=0.53]{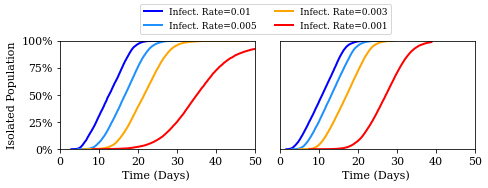}
        \begin{minipage}{0.55\columnwidth}
	\centering
	\footnotesize
	(a) Population = 10,000.
\end{minipage}
    \begin{minipage}{0.42\columnwidth}
	\centering
	\footnotesize
	(b) Population = 20,000.
\end{minipage}

        % \subfloat[Population = 10,000.]{\label{subfig:isolated-10k}\includegraphics[scale=0.43]{figs/isolated_pop10000.pdf}}
        % \hfill
        % \subfloat[Population = 20,000.]{\label{subfig:isolated-20k}\includegraphics[scale=0.43]{figs/isolated_pop20000.pdf}}
    \caption{Percentage of isolated population.}
    \label{fig:symptom-simulation-results-isolated}
\end{figure}

As a companion to Fig.~\ref{fig:symptom-simulation-results}, we also present the portions of ``active while infected" and isolated agents, see Fig.~\ref{fig:symptom-simulation-results-active}.
In Fig.~\ref{fig:symptom-simulation-results-active},
% to improve the visibility of results, we only show the 
% percentage of ``active while infected'' patients in the 10,000 population with the two extreme infection rates: 0.001 and 0.01
the benefit of patient isolation can be seen clearly: the percentage of active infected population is decreasing after showing a peak, which limits the speed of the spread of the disease.
The percentage of isolated population shown in Fig.~\ref{fig:symptom-simulation-results-isolated}  explains the decrease of active infected population.
After more agents show symptoms and are isolated, the active infected population starts dropping. 

%\lishan{[Future work, Not sure here is the best place to say this. Maybe in conclusion.] In this paper we only show the scenario of applying patient isolation, but our simulation tool can be extended to support more scenarios. We list this as future work.}

% \begin{figure*}[tb]
% 	\centering
% 	\begin{minipage}[t]{\textwidth}
% 		\centering
%  		\includegraphics[scale=0.8]{figs/patient_states.png}
% 	\end{minipage}
% % 	\vspace{-\baselineskip}
% % 	\vspace{-1em}
% 	\caption{State transition of agents using different distributions in Simulation Basic+Symptom.}
% 	\label{fig:patient_states}
% % 	\vspace{-1em}
% \end{figure*}

\rev{\section{Discussion and Limitations}
\label{sec:discussion}

The proposed approach allows investigating the spreading of COVID-19 in an urban population.
Incomplete and missing data may limit its generalization and make it far from being the definitive COVID-19 spreading model.
Main limitations of our approach are discussed in this section.

\noindent \textbf{Scarcity of data.}
We continue to seek additional data sets on COVID-19 outbreaks to increase the accuracy of simulations, however, the current lack of data is an unfortunate limitation.
The limited amount of data for super spreaders affects observations on their mobility, see Fig.~\ref{fig:mobility}\subref{subfig:mobility_spread}.
Selfish decisions (e.g., people that ignore mask mandate) are considered in Section \ref{sec:simulation} by using different infect ratios due to missing information about protective measures adopted by patients.
The infect ratio increases with the number of selfish agents.
Our approach accounts for irresponsible behaviors (i.e., people moving after being infected) as presented in Section \ref{sec:irresponsible}.
We overcome the absence of data about movements inside buildings (i.e., rooms and floors) by reducing the population size.

\noindent \textbf{Privacy concerns.}
The KCDC data set is anonymized and no sensitive data of monitored patients can be retrieved.
No data about the underage population is provided as well as movements of patients from/to their private homes.
This limits scenarios that can be analyzed, e.g., the impact of school closures or the spreading of the virus in households.

\noindent \textbf{Input parameter assumptions.}
The KCDC data set does not show the mode of transport.
We overcome this limitation by assuming a pedestrian:vehicles ratio of 1:1.
Input parameters can be fully customized and other researchers using our approach can easily change these values.}
\section{Related Work}
\label{sec:related}

Agent-based models (ABMs) are used in the literature as an alternative to mathematical models~\cite{bithell2008discrete} to study interactions of individuals and investigate their impact on the considered system~\cite{kelly2013selecting}.
ABMs are used for different purposes, such as modeling pedestrian movements or resource exploitation~\cite{o2012agent}.
ABMs have been also used for studying the spread of diseases (e.g., cholera~\cite{crooks2014agent} and Ebola~\cite{jacobsen2016lessons,venkatramanan2018using}) by modeling interactions between humans and their environment.
ABMs are often used with geographical information systems (GIS) to simulate interactions of individuals in an urban context and investigate the spread of a diseases in a given area~\cite{wang2010use}.

Ferguson et al.~\cite{ferguson2006strategies} simulate and study the spread of influenza in British and American households, schools, and workplaces. They use different data sets to parameterize their simulation, such as population density data and travel patterns. The impact of different containment strategies (e.g., travel restrictions or vaccination) is evaluated against the \textit{attack rate} (i.e., $R_0$, the percentage of at risk population that is infected by the virus).
\rev{Only large scale (international) movements are considered in \cite{ferguson2006strategies} and it is observed that stricter border controls delay the peak of the disease by several days. Instead, we define mobility to account for movements in urban areas.}
%Interactions of individuals with each other are not considered.

The COVID-19 pandemic has been largely studied in recent months due to its disruptive effects on public health.
Bi et al.~\cite{bi2020epidemiology} use a model based on conditional logistic regression to study the transmission of COVID-19 in Shenzhen, China.
Using data from contact-based surveillance and accurate infector-infectee relationships, they confirm that, on average, COVID-19 has a short incubation period (i.e., less than a week) and a long clinical course.
Garg et al.~\cite{garg2020hospitalization} present a framework to predict hospitalization rates from clinical data (e.g., age, ethnicity, medical conditions, clinical course) of COVID-19 patients in 14 states of the USA.
Pung et al.~\cite{pung2020investigation} collect epidemiological and clinical data from COVID-19 patients to study the spreading of the virus in three different Singapore clusters.
\rev{They account for agent mobility by analyzing cluster interactions that facilitate the spread of the virus. Differently from~\cite{pung2020investigation}, we define agent mobility as the probability that agents leave the building where they are.}
%They
%also analyzed people's interactions and the virus transmission and
%conclude that contact tracing is crucial for limiting the spread of COVID-19.

Bock et al.~\cite{bock2020mitigation} develop a simulation model that investigates the efficiency of mitigation strategies in slowing down the spread of the virus by accounting for interactions within households (where it is hard to social distance).
They use census data and age distribution of Germany and Poland as input parameters of their model.
\rev{Pej{\'o}  and Bicz{\'o}k~\cite{pejo2020corona} use game theory to evaluate the efficiency of face masks and social distancing in limiting the spread of COVID-19 when there are selfish agents that do not use any countermeasures.
Similarly, Bhattacharyya and Bauch~\cite{bhattacharyya2011wait} use game theory to evaluate the efficiency of protective vaccines (the safest way to achieve herd immunity \cite{d2020s}) when selfish agents refuse to get the shot.}
Rader et al.~\cite{rader2020crowding} study the spread of COVID-19 using spatial, urbanization, climate, and census data of worldwide cities.
They observe that cities that are more densely populated are affected by larger incidence and more prolonged epidemic.

Grossmann et al.~\cite{grossmann2020importance} propose a stochastic network-based COVID-19 spreading model and compare its results with those obtained through an ordinary differential equations (ODE) model.
In their network-based model, they use random graph models to represent interaction structures and human connections.
They observe that ODE models struggle to correctly represent inhomogeneity of interaction structures, a feature that profoundly affects the spread of the virus.

Rockett et al.~\cite{rockett2020revealing} use ABM parameterized with census data and transmission pathways to investigate the spread of COVID-19 in Australia.
Kim et al.~\cite{kim2020location} use synthetic location-based social network data sets to study outbreak diseases (e.g., COVID-19) and evaluate the effectiveness of different mitigation strategies.
None of these studies extract people movements and dynamics from available data sets to parameterize models and analyze the spread of the virus.
\vspace{-1mm}

\section{Conclusions}
\label{sec:conclusion}

Information and routes of South Korean COVID-19 patients are analyzed  to study the disease outbreak in the Gangnam district of Seoul.
Movement habits in South Korea are extracted by available data sets to parameterize simulations, based on ABM and GIS, and study interactions among people.
Preliminary results show that the proposed approach correctly associates the virus spread velocity with the virus infectiousness and the population size.

This GeoMason model can be used to flexibly examine and evaluate a wide variety of different scenarios based on patterns in real-world data. While it is not a definitive COVID-19 spread model, it can be used to investigate useful \textit{what-if} scenarios.
We are currently working on expanding the simulation model to evaluate the effectiveness of partial lockdown measures such as placing distance restrictions and curfews.

% In this paper, we analyze data of South Korean COVID-19 patient and their routes collected by the KCDC.
% We extract information about movement habits in South Korea and use this data to parameterize simulations based on ABM and GIS to study people interactions in Seoul area.
% Preliminary results show that \ricc{add some comments about observed results}.
% Observations presented in this paper may help researchers to investigate real-world scenarios and provide decision makers with results based on real data to stop the COVID-19 outbreak.
% In the future, \ricc{add future work}.

\bibliographystyle{IEEEtran}
\bibliography{REF}

\begin{thebibliography}{10}
\providecommand{\url}[1]{#1}
\csname url@samestyle\endcsname
\providecommand{\newblock}{\relax}
\providecommand{\bibinfo}[2]{#2}
\providecommand{\BIBentrySTDinterwordspacing}{\spaceskip=0pt\relax}
\providecommand{\BIBentryALTinterwordstretchfactor}{4}
\providecommand{\BIBentryALTinterwordspacing}{\spaceskip=\fontdimen2\font plus
\BIBentryALTinterwordstretchfactor\fontdimen3\font minus
  \fontdimen4\font\relax}
\providecommand{\BIBforeignlanguage}[2]{{%
\expandafter\ifx\csname l@#1\endcsname\relax
\typeout{** WARNING: IEEEtran.bst: No hyphenation pattern has been}%
\typeout{** loaded for the language `#1'. Using the pattern for}%
\typeout{** the default language instead.}%
\else
\language=\csname l@#1\endcsname
\fi
#2}}
\providecommand{\BIBdecl}{\relax}
\BIBdecl

\bibitem{rothan2020epidemiology}
H.~A. Rothan and S.~N. Byrareddy, ``The epidemiology and pathogenesis of
  coronavirus disease (covid-19) outbreak,'' \emph{Journal of autoimmunity}, p.
  102433, 2020.

\bibitem{who2020pandemic}
``Who director-general's opening remarks at the media briefing on covid-19 --
  11 march 2020,''
  \url{shorturl.at/crAMW},
  2020, [Online; accessed 23-September-2020].

\bibitem{feng2020rational}
S.~Feng, C.~Shen, N.~Xia, W.~Song, M.~Fan, and B.~J. Cowling, ``Rational use of
  face masks in the covid-19 pandemic,'' \emph{The Lancet Respiratory
  Medicine}, vol.~8, no.~5, pp. 434--436, 2020.

\bibitem{lewnard2020scientific}
J.~A. Lewnard and N.~C. Lo, ``Scientific and ethical basis for
  social-distancing interventions against covid-19,'' \emph{The Lancet.
  Infectious diseases}, vol.~20, no.~6, p. 631, 2020.

\bibitem{sjodin2020only}
H.~Sj{\"o}din, A.~Wilder-Smith, S.~Osman, Z.~Farooq, and J.~Rockl{\"o}v, ``Only
  strict quarantine measures can curb the coronavirus disease (covid-19)
  outbreak in italy, 2020,'' \emph{Eurosurveillance}, vol.~25, no.~13, 2020.

\bibitem{nussbaumer2020quarantine}
B.~Nussbaumer-Streit, V.~Mayr, A.~I. Dobrescu, A.~Chapman, E.~Persad,
  I.~Klerings, G.~Wagner, U.~Siebert, D.~Ledinger, C.~Zachariah \emph{et~al.},
  ``Quarantine alone or in combination with other public health measures to
  control covid-19: a rapid review,'' \emph{Cochrane Database of Systematic
  Reviews}, no.~9, 2020.

\bibitem{saez2020effectiveness}
M.~Saez, A.~Tobias, D.~Varga, and M.~A. Barcel{\'o}, ``Effectiveness of the
  measures to flatten the epidemic curve of covid-19. the case of spain,''
  \emph{Science of the Total Environment}, p. 138761, 2020.

\bibitem{matrajt2020evaluating}
L.~Matrajt and T.~Leung, ``Evaluating the effectiveness of social distancing
  interventions to delay or flatten the epidemic curve of coronavirus
  disease,'' \emph{Emerging infectious diseases}, vol.~26, no.~8, p. 1740,
  2020.

\bibitem{smetters2020stay}
K.~A. Smetters, ``Stay-at-home orders and second waves: a graphical
  exposition,'' \emph{The Geneva Risk and Insurance Review}, pp. 1--10, 2020.

\bibitem{who2020lifting}
``Who director-general's opening remarks at the media briefing on covid-19 --
  10 april 2020,''
  \url{shorturl.at/fquDW},
  2020, [Online; accessed 23-September-2020].

\bibitem{kim2020location}
J.-S. Kim, H.~Kavak, C.~O. Rouly, H.~Jin, A.~Crooks, D.~Pfoser, C.~Wenk, and
  A.~Z{\"u}fle, ``Location-based social simulation for prescriptive analytics
  of disease spread,'' \emph{SIGSPATIAL Special}, vol.~12, no.~1, pp. 53--61,
  2020.

\bibitem{bock2020mitigation}
W.~Bock, B.~Adamik, M.~Bawiec, V.~Bezborodov, M.~Bodych, J.~P. Burgard,
  T.~Goetz, T.~Krueger, A.~Migalska, B.~Pabjan \emph{et~al.}, ``Mitigation and
  herd immunity strategy for covid-19 is likely to fail,'' \emph{medRxiv},
  2020.

\bibitem{kcdc2020covid19}
K.~C. for Disease Control \&~Prevention, ``Coronavirus disease-19, republic of
  korea,'' \url{http://ncov.mohw.go.kr/en/}, 2020, [Online; accessed
  17-September-2020].

\bibitem{googlemap}
``Google maps,'' \url{https://www.google.com/maps/}, 2020, [Online; accessed
  17-September-2020].

\bibitem{kakaomap}
``Kakao map,'' \url{https://map.kakao.com/}, 2020, [Online; accessed
  17-September-2020].

\bibitem{navermap}
``Naver map,'' \url{https://m.map.naver.com/}, 2020, [Online; accessed
  17-September-2020].

\bibitem{kim2020ds4c}
J.~Kim and J.~Lee, ``Data science for covid-19 (ds4c),''
  \url{https://www.kaggle.com/kimjihoo/coronavirusdataset}, 2020, [Accessed on
  2020-09-25].

\bibitem{kim2020spatiotemporal}
S.~Kim and M.~C. Castro, ``Spatiotemporal pattern of covid-19 and government
  response in south korea (as of may 31, 2020),'' \emph{International Journal
  of Infectious Diseases}, vol.~98, pp. 328--333, 2020.

\bibitem{sullivan2010geomason}
K.~Sullivan, M.~Coletti, and S.~Luke, ``Geomason: Geospatial support for
  mason,'' Department of Computer Science, George Mason University, Tech. Rep.,
  2010.

\bibitem{crooks2014agent}
A.~Crooks and A.~Hailegiorgis, ``An agent-based modeling approach applied to
  the spread of cholera,'' \emph{Environmental Modelling \& Software}, vol.~62,
  pp. 164--177, 2014.

\bibitem{openstreetmap}
``Openstreetmap,'' \url{https://www.openstreetmap.org/}, 2020, [Online;
  accessed 17-September-2020].

\bibitem{CDCguidance}
CDC, ``Public health guidance for community-related exposure,''
  \url{https://www.cdc.gov/coronavirus/2019-ncov/php/public-health-recommendations.html},
  2020, [Accessed on 2020-10-14].

\bibitem{whoQandA}
WHO, ``Q\&a on coronaviruses (covid-19),''
  \url{https://www.who.int/emergencies/diseases/novel-coronavirus-2019/question-and-answers-hub/q-a-detail/q-a-coronaviruses},
  2020, [Accessed on 2020-10-09].

\bibitem{seoulGIS}
``Osm extracts for seoul,''
  \url{https://download.bbbike.org/osm/bbbike/Seoul/}, 2020, [Accessed on
  2020-10-09].

\bibitem{bithell2008discrete}
M.~Bithell, J.~Brasington, and K.~Richards, ``Discrete-element,
  individual-based and agent-based models: Tools for interdisciplinary enquiry
  in geography?'' \emph{Geoforum}, vol.~39, no.~2, pp. 625--642, 2008.

\bibitem{kelly2013selecting}
R.~A. Kelly, A.~J. Jakeman, O.~Barreteau, M.~E. Borsuk, S.~ElSawah, S.~H.
  Hamilton, H.~J. Henriksen, S.~Kuikka, H.~R. Maier, A.~E. Rizzoli
  \emph{et~al.}, ``Selecting among five common modelling approaches for
  integrated environmental assessment and management,'' \emph{Environmental
  modelling \& software}, vol.~47, pp. 159--181, 2013.

\bibitem{o2012agent}
D.~O’Sullivan, J.~Millington, G.~Perry, and J.~Wainwright, ``Agent-based
  models--because they’re worth it?'' in \emph{Agent-based models of
  geographical systems}.\hskip 1em plus 0.5em minus 0.4em\relax Springer, 2012,
  pp. 109--123.

\bibitem{jacobsen2016lessons}
K.~H. Jacobsen, A.~A. Aguirre, C.~L. Bailey, A.~V. Baranova, A.~T. Crooks,
  A.~Croitoru, P.~L. Delamater, J.~Gupta, K.~Kehn-Hall, A.~Narayanan
  \emph{et~al.}, ``Lessons from the ebola outbreak: action items for emerging
  infectious disease preparedness and response,'' \emph{EcoHealth}, vol.~13,
  no.~1, pp. 200--212, 2016.

\bibitem{venkatramanan2018using}
S.~Venkatramanan, B.~Lewis, J.~Chen, D.~Higdon, A.~Vullikanti, and M.~Marathe,
  ``Using data-driven agent-based models for forecasting emerging infectious
  diseases,'' \emph{Epidemics}, vol.~22, pp. 43--49, 2018.

\bibitem{wang2010use}
J.~Wang, J.~Xiong, K.~Yang, S.~Peng, and Q.~Xu, ``Use of gis and agent-based
  modeling to simulate the spread of influenza,'' in \emph{2010 18th
  International Conference on Geoinformatics}.\hskip 1em plus 0.5em minus
  0.4em\relax IEEE, 2010, pp. 1--6.

\bibitem{ferguson2006strategies}
N.~M. Ferguson, D.~A. Cummings, C.~Fraser, J.~C. Cajka, P.~C. Cooley, and D.~S.
  Burke, ``Strategies for mitigating an influenza pandemic,'' \emph{Nature},
  vol. 442, no. 7101, pp. 448--452, 2006.

\bibitem{bi2020epidemiology}
Q.~Bi, Y.~Wu, S.~Mei, C.~Ye, X.~Zou, Z.~Zhang, X.~Liu, L.~Wei, S.~A. Truelove,
  T.~Zhang \emph{et~al.}, ``Epidemiology and transmission of covid-19 in
  shenzhen china: Analysis of 391 cases and 1,286 of their close contacts,''
  \emph{MedRxiv}, 2020.

\bibitem{garg2020hospitalization}
S.~Garg, ``Hospitalization rates and characteristics of patients hospitalized
  with laboratory-confirmed coronavirus disease 2019—covid-net, 14 states,
  march 1--30, 2020,'' \emph{MMWR. Morbidity and mortality weekly report},
  vol.~69, 2020.

\bibitem{pung2020investigation}
R.~Pung, C.~J. Chiew, B.~E. Young, S.~Chin, M.~I. Chen, H.~E. Clapham, A.~R.
  Cook, S.~Maurer-Stroh, M.~P. Toh, C.~Poh \emph{et~al.}, ``Investigation of
  three clusters of covid-19 in singapore: implications for surveillance and
  response measures,'' \emph{The Lancet}, 2020.

\bibitem{pejo2020corona}
B.~Pej{\'o} and G.~Bicz{\'o}k, ``Corona games: Masks, social distancing and
  mechanism design,'' in \emph{Proceedings of the 1st ACM SIGSPATIAL
  International Workshop on Modeling and Understanding the Spread of COVID-19},
  2020, pp. 24--31.

\bibitem{bhattacharyya2011wait}
S.~Bhattacharyya and C.~Bauch, ``“wait and see” vaccinating behaviour
  during a pandemic: a game theoretic analysis,'' \emph{Vaccine}, vol.~29,
  no.~33, pp. 5519--5525, 2011.

\bibitem{d2020s}
G.~D’Souza and D.~Dowdy, ``What’s herd immunity and how can we achieve it
  with covid-19,''
  \url{https://www.jhsph.edu/covid-19/articles/achieving-herd-immunity-with-covid19.html},
  2020, [Accessed on 2020-11-30].

\bibitem{rader2020crowding}
B.~Rader, S.~Scarpino, A.~Nande, A.~Hill, R.~Reiner, D.~Pigott, B.~Gutierrez,
  M.~Shrestha, J.~Brownstein, M.~Castro \emph{et~al.}, ``Crowding and the
  epidemic intensity of covid-19 transmission,'' \emph{medRxiv}, 2020.

\bibitem{grossmann2020importance}
G.~Grossmann, M.~Backenkoehler, and V.~Wolf, ``Importance of interaction
  structure and stochasticity for epidemic spreading: A covid-19 case study,''
  \emph{medRxiv}, 2020.

\bibitem{rockett2020revealing}
R.~J. Rockett, A.~Arnott, C.~Lam, R.~Sadsad, V.~Timms, K.-A. Gray, J.-S. Eden,
  S.~Chang, M.~Gall, J.~Draper \emph{et~al.}, ``Revealing covid-19 transmission
  in australia by sars-cov-2 genome sequencing and agent-based modeling,''
  \emph{Nature medicine}, pp. 1--7, 2020.

\end{thebibliography}
% {\small
% \printbibliography}

\end{document}